\documentclass[sort]{elsarticle}

\usepackage{lineno}
\usepackage[hidelinks]{hyperref}
\modulolinenumbers[5]       
\usepackage[english]{babel}
\newtheorem{assumption}{\textbf{Assumption}}
\newtheorem{lemma}{\textbf{Lemma}}
\newtheorem{definition}{\textbf{Definition}}
\newtheorem{theorem}{\textbf{Theorem}}

\newtheorem{remark}{\textbf{Remark}}

\newtheorem{problem}{\textbf{Problem}}

\IfFileExists{newtxmath.sty}
{
\usepackage{newtxtext,newtxmath}
}{
\usepackage{mathptmx}
}
\usepackage{times}
\usepackage{amsmath}
\usepackage{amsfonts}
\usepackage{mathrsfs}
\usepackage{xcolor}
\usepackage{graphicx}
\usepackage{varioref}
\usepackage{placeins}
\usepackage[breakable]{tcolorbox}

\newcommand{\T}{^{\mbox{\tiny T}}}
\newcommand{\R}{\mathbb{R}}

\let\leq\leqslant
\let\geq\geqslant

\newenvironment{proof}[1][Proof]%
{\par\noindent\textit{#1:\ }}%
{\hspace*{\fill} \rule{6pt}{6pt}}
\newenvironment{proof*}[1][Proof]%
{\par\noindent\textit{#1:\ }}{}

\DeclareMathOperator{\im}{Im}

\DeclareMathOperator{\Ker}{Ker}
\DeclareMathOperator{\Span}{span}
\usepackage{tikz}
\usetikzlibrary{shapes,calc,arrows,patterns,decorations.pathmorphing
	,decorations.markings}
\usetikzlibrary{arrows.meta}

\newenvironment{system}[1]%
{\setlength{\arraycolsep}{0.5mm}\left\{ \; \begin{array}{#1}}%
	{\end{array} \right.}
\newenvironment{system*}[1]%
{\setlength{\arraycolsep}{0.5mm} \begin{array}{#1}}%
	{\end{array}}

\begin{document}

\begin{frontmatter}
  \title{Scale-free weak output synchronization of
    multi-agent systems with adaptive protocols} 
  
  \author[Netherlands]{Anton A. Stoorvogel}\ead{A.A.Stoorvogel@utwente.nl}
  \author[WSU1]{Ali Saberi}\ead{saberi@wsu.edu}
  \author[NEU]{Zhenwei Liu\texorpdfstring{\corref{myc}}{}}
  \cortext[myc]{Corresponding author. This work is supported by the
    National Natural Science Foundation of China under Grant 62273084
    and the Liaoning Provincial International Science and Technology
    Cooperation Program under Grant
    2025JH2/101900030.}\ead{liuzhenwei@ise.neu.edu.cn}
  \author[NEU]{Qiaofeng Wen}\ead{wenqiaofeng@stumail.neu.edu.cn}
  \address[Netherlands]{Department of Electrical Engineering,
    Mathematics and Computer Science, University of Twente, Enschede,
    The Netherlands} \address[WSU1]{School of Electrical Engineering
    and Computer Science, Washington State University, Pullman WA,
    USA} \address[NEU]{College of Information Science and Engineering,
    Northeastern University, Shenyang, P. R. China}
  
  \begin{abstract}
    In this paper, we study output synchronization for multi-agent
    systems. The objective is to design a protocol which only depends
    on the agent dynamics and does not require any knowledge of the
    network. If the network has a directed spanning tree then the
    protocols designed in this paper achieve classical output
    synchronization. Otherwise, the protocol achieves weak
    synchronization which is induced by network stability in the sense
    that the signals exchanged over the network converge to zero. Weak
    sychronization is explained in detail in this paper. Even though
    we consider linear agents, it is known that this in general
    requires nonlinear protocols. In the paper we use adaptive
    protocols. In the literature, two classes of protocols are
    considered often called collaborative protocols (with additional
    communication between the protocols and non-collaborative protocols
    (sometimes referred to as fully decentralized where the additional
    communication is not present). This paper considers both of these
    cases.
  \end{abstract}


\end{frontmatter}

\section{Introduction}

Over the past two decades, distributed control of multi-agent systems
(MAS) have been extensively studied, see
\cite{ren-atkins,saber-murray2, wu-chua2}.  In general, the MAS is
consisting of many components with a constrained or limited
communication network which is similar to the system in decentralized
control research \cite{siljak,corfmat-morse2}. Applications include
networks of generators interconnected via power grids and vehicular
platoons in transportation systems. Early decentralized control
architectures often relied on a designated supervisory agent to
coordinate communication with other agents. However, this design
introduced a single point of failure, making the system highly
vulnerable to disruptions in the network.

Classically protocol design required knowledge of both agents and the
network. However, it is in many applications impossible to obtain
complete knowledge of the network. Later work limited the required
knowledge about the network by, for instance, only making assumptions
such as lower or upper bounds on the spectrum of the Laplacian matrix
associated to the graph describing the network structure. But even
these bounds are hard to obtain.  In recent years, adaptive nonlinear
protocols were developed to achieve state synchronization for MAS, see
\cite{li-wen-duan-ren-tac-2015,lv-li-ren-duan-chen-auto-2016,liu-zhang-saberi-stoorvogel-auto,%
  lv-li-scis-2021}. These protocols do not need assumptions, like a
bound on the Laplacian matrix's non-zero eigenvalues, by using an
adaptive time-varying protocol gain. However, it still requires that
the network is strongly connected or has a direct spanning tree. This
actually still inherently has some of the difficulties presented
before. How can we check if this connectivity is present in the
network? Secondly, what happens if the network fail this assumption?

Recently, the concept of weak synchronization was proposed in
\cite{stoorvogel-saberi-liu-weak}. This basically means that our
protocol achieves network stability, all signals exchanged over the
network converge to zero. If the network happens to have a directed
spanning tree then we will have classical output
synchronization. However, in general we obtain weak synchronization
which implies synchronization within basic bicomponents. The output of
agents which are not part of a basic bicomponent converge to a convex
combination of the outputs of the agents contained in the basic
bicomponents. Moverover, this convergence structure is completely determined by
the network structure and independent of the specific protocol or
initial conditions.

In this paper, we design adaptive nonlinear protocol which achieve
weak output synchronization for any network without any assumptions on
the network. We design two kinds of protocols, non-collaborative and
collaborative.  If the network happens to have a directed spanning
tree then we obtain classical output synchronization. Meanwhile, if
this is not the case, then weak output synchronization can be
achieved.

There are three major innovations compared to earlier work.
\begin{itemize}
\item In this paper we look at output (weak) synchronization where
  ouputs and not states are communicated via the network graph. This
  implies the need for decentralized observer design while previous
  papers using adaptive protocols were designed for state
  synchronization which is a much simpler case.
\item Compared to earlier linear protocol design such as
  \cite{stoorvogel-saberi-liu2}, the adaptive design enables us to
  drop restrictions on the agent model such as restrictions on the system poles.
\item The protocols designed in this paper are completely independent
  of the network structure. In particular, it does not require any
  connectivity assumptions.
\end{itemize}

\subsection*{Preliminaries on graph theory}

Given a matrix $M\in \mathbb{R}^{m\times n}$, $M\T$ denotes its
conjugate transpose. A square matrix $M$ is said to be Hurwitz stable
if all its eigenvalues are in the open left half complex plane.
$\im M$ denotes the image of matrix $M$. $M_a\otimes M_b$ depicts the
Kronecker product of $M_a$ and $M_b$. $I_n$ denotes the
$n$-dimensional identity matrix and $0_n$ denotes $n\times n$ zero
matrix; sometimes we drop the subscript if the dimension is clear from
the context. For a signal $u$, we denote the $L_2$ norm by $\|u\|$ or
$\|u\|_2$, and the $L_\infty$ norm by $\|u\|_\infty$.

To describe the information flow among the agents we associate a
\emph{weighted graph} $\mathcal{G}$ to the communication network. The
weighted graph $\mathcal{G}$ is defined by a triple
$(\mathcal{V}, \mathcal{E}, \mathcal{A})$ where
$\mathcal{V}=\{1,\ldots, N\}$ is a node set, $\mathcal{E}$ is a set of
pairs of nodes indicating connections among nodes, and
$\mathcal{A}=[a_{ij}]\in \mathbb{R}^{N\times N}$ is the weighted
adjacency matrix with non negative elements $a_{ij}$. Each pair in
$\mathcal{E}$ is called an \emph{edge}, where $a_{ij}>0$ denotes an
edge $(j,i)\in \mathcal{E}$ from node $j$ to node $i$ with weight
$a_{ij}$. Moreover, $a_{ij}=0$ if there is no edge from node $j$ to
node $i$. We assume there are no self-loops, i.e.\ we have
$a_{ii}=0$. A \emph{path} from node $i_1$ to $i_k$ is a sequence of
nodes $\{i_1,\ldots, i_k\}$ such that $(i_j, i_{j+1})\in \mathcal{E}$
for $j=1,\ldots, k-1$. A \emph{directed tree} is a subgraph (a subset of
nodes and edges) in which every node has exactly one parent node
except for one node, called the \emph{root}, which has no parent
node. A \emph{directed spanning tree} is a subgraph which is a
directed tree containing all the nodes of the original graph. If a
directed spanning tree exists, the root has a directed path to every
other node in the tree \cite{royle-godsil}.

For a weighted graph $\mathcal{G}$, the matrix
$L=[\ell_{ij}]$ with
\[
  \ell_{ij}=
  \begin{system}{cl}
    \sum_{k=1}^{N} a_{ik}, & i=j,\\
    -a_{ij}, & i\neq j,
  \end{system}
\]
is called the \emph{Laplacian matrix} associated with the graph
$\mathcal{G}$. The Laplacian matrix $L$ has all its eigenvalues in the
closed right half plane and at least one eigenvalue at zero associated
with right eigenvector $\textbf{1}$ (a vector with all its entries
equal to $1$), see \cite{royle-godsil}. Moreover, if
the graph contains a directed spanning tree, the Laplacian matrix $L$
has a single eigenvalue at the origin and all other eigenvalues are
located in the open right-half complex plane \cite{ren-book}.

In the absence of a directed spanning tree, the Laplacian matrix of
the graph has an eigenvalue at the origin with a multiplicity
$k>1$. This implies that it is a $k$-reducible matrix and the graph
has $k$ basic bicomponents.  The book \cite[Definition 2.19]{wu-book}
shows that, after a suitable permutation of the nodes, a Laplacian
matrix with $k$ basic bicomponents can be written in the following
form:
\begin{equation}\label{Lstruc}
  L=\begin{pmatrix}
    L_0 & L_{01}     & L_{02} & \cdots  & L_{0k} \\
    0   & L_1        & 0      & \cdots  & 0 \\
    \vdots & \ddots  & \ddots & \ddots  & \vdots \\
    \vdots &         & \ddots & L_{k-1} & 0 \\
    0      & \cdots & \cdots  & 0       & L_k
  \end{pmatrix}
\end{equation}
where $L_1,\ldots, L_k$ are the Laplacian matrices associated to the
$k$ basic bicomponents $\{ \mathcal{B}_1,\ldots, \mathcal{B}_k\}$ in
our network. These matrices have a simple eigenvalue in $0$ because
they are associated with a strongly connected component. On the other
hand, $L_0$ contains all non-basic bicomponents and is a grounded
Laplacian with all eigenvalues in the open right-half plane. After
all, if $L_0$ would be singular then the network would have an
additional basic bicomponent.

\section{Weak synchronization of MAS}

In this section, we introduce the concept of weak synchronization for
homogeneous MAS. Consider $N$ homogeneous agents
\begin{equation}\label{system}
  \begin{system*}{cl}
    \dot{x}_i &= Ax_i +B u_i,  \\
    y_i &= Cx_i,
  \end{system*}
\end{equation}
where $x_i\in\mathbb{R}^{n}$, $u_i\in\mathbb{R}^{m}$ and
$y_i\in\mathbb{R}^{p}$ are the state, input, output of agent $i$ for
$i=1,\ldots, N$ with $(A,B)$ stabilizable and $(C,A)$ detectable.

The communication network provides agent $i$ with the following
information which is a linear combination of its own output relative
to that of other agents
\begin{equation}\label{zeta1}
  \zeta_i=\sum_{j=1}^{N}a_{ij}(y_i-y_j),
\end{equation}
where $a_{ij}\geq 0$ and $a_{ii}=0$. The communication topology of the
network can be described by a weighted and directed graph
$\mathcal{G}$ with nodes corresponding to the agents in the network
and the weight of edges given by coefficient $a_{ij}$. We denote by
$\mathbb{G}^N$ the set of all graphs with $N$ nodes. In terms of the
coefficients of the associated Laplacian matrix $L$, $\zeta_i$ can be
rewritten as
\begin{equation}\label{zeta}
  \zeta_i= \sum_{j=1}^{N}\ell_{ij}y_j.
\end{equation}
Our protocols are of the form:
\begin{equation}\label{protocoln}
  \begin{system*}{ccl}
    \dot{\xi}_i &=& f_i(\xi_i, \zeta_i) \\
    u_i &=& g_i(\xi_i, \zeta_i)
  \end{system*}
\end{equation}
These protocols are referred to as non-collaborative protocols to
distinguish them from collaborative protocols defined later in this
paper.

In the following, we introduce the concept of network stability which
yields weak synchronization which is vastly different from output synchronization.

\begin{definition}[\textbf{Non-collaborative network stability}]
  Consider a multi-agent network described by \eqref{system} and
  \eqref{zeta}. A non-collaborative protocol of the form
  \eqref{protocoln} achieves network stability if:
  \[
    \zeta_i(t)=\sum_{j=1}^{N}a_{ij}(y_i-y_j)\to 0
  \]
  as $t\to \infty$, for any $i \in \{1,\ldots, N\}$ and for all
  possible initial conditions
\end{definition}

We are considering output synchronization in this paper. In
\cite{li-duan-chen-huang}, an extra communication between the
protocols was introduced which significantly simplified the required
observer design. Protocols wich use this extra information are
referred to as collaborative protocols. To be more specific, we assume
there is an additional communication possible over the same network
between the protocols of the different agents. In particular, for the
collaborative protocol of agent $i$, besides \eqref{zeta}, they can
also use
\begin{equation}\label{zetau}
	\tilde{\zeta}_i= \sum_{j=1}^{N}\ell_{ij} x_{i,c}
\end{equation}
which communicates the state of the protocol of the different agents
over the network. This extra communication significantly reduces the
number of assumptions we need to make on our agent model. Collaborative
protocols are of the form:
\begin{equation}\label{out_dyncol}
  \begin{system*}{ccl}
     \dot{x}_{i,c}&=f(x_{i,c},\zeta_i,\tilde{\zeta}_i),\\
      u_i&=g(x_{i,c},\zeta_i,\tilde{\zeta}_i),
  \end{system*}
\end{equation}

\begin{definition}[\textbf{Collaborative network stability}]\label{delta-level2}
  Consider a multi-agent network described by \eqref{system} and
  \eqref{zeta}. A collaborative protocol of the form
  \eqref{protocoln} achieves network stability if:
  \[
    \zeta_i(t)=\sum_{j=1}^{N}a_{ij}(y_i-y_j)\to 0
  \]
  and
  \[
    \hat{\zeta}_i(t)=\sum_{j=1}^{N}a_{ij}(x_{i,c}-x_{j,c})\to 0
  \]
  as $t\to \infty$, for any $i \in \{1,\ldots, N\}$ and for all
  possible initial conditions
\end{definition}

\begin{lemma}\label{xxxxx2}
  Consider an MAS described by \eqref{system} and
  \eqref{zeta1}. Assume a non-collaborative protocol \eqref{protocoln}
  or a collaborative protocol \eqref{out_dyncol} achieves network
  stability. In either case, we achieve what is referred to as weak
  synchronization, i.e.
  \begin{itemize}
  \item If the network contains a directed spanning tree then we
    achieve output synchronization.
  \item If the network does not have a directed spanning tree
    which implies that the graph has $k>1$ basic bicomponents then:
    \begin{itemize}
    \item Within basic bicomponent $i$ we have output synchronization
      in the sense that there exists a signal $y^i_s$ such that
      $y_j(t)-y^i_s(t)\rightarrow 0$ for any agent $j$ which is part of
      basic bicomponent $i$.
    \item An agent $j$ which is not part of any of the basic
      bicomponents synchronizes to a trajectory $y_{j,s}$,
      \begin{equation}\label{ys}
        y_{j,s}=\sum_{i=1}^k\, \beta_{j,i} y^i_s 
      \end{equation}
      where the coefficients $\beta_{j,i}$ are nonnegative, satisfy:
      \begin{equation}\label{betasum}
        1=\sum_{i=1}^k\, \beta_{j,i} 
      \end{equation}
      and only depend on the parameters of the network and do not depend
      on any of the initial conditions.
    \end{itemize}
  \end{itemize}
\end{lemma}

\begin{proof}
  The proof of this lemma is a minor adaptation of the proof of the
  same result in \cite{stoorvogel-saberi-liu-weak} to include
  nonlinear protocols.
\end{proof}

In the above, we have seen that a collaborative or non-collaborative
protocol achieves the very natural concept of weak synchronization
when it achieves network stability. The main part of this paper is
about the design of adaptive protocols that achieve weak
synchronization. The design is split into the two cases of
non-collaborative and collaborative protocols as defined above.

\begin{problem}\label{prob_x}
  Consider a MAS \eqref{system} with associated network communication
  \eqref{zeta}. The \textbf{scale-free non-collaborative weak output
    synchronization problem} is to find, if possible, a
  non-collaborative nonlinear non-collaborative protocol of the form
  \eqref{protocoln} using only knowledge of agent models, i.e.,
  $(A, B,C)$, such that the MAS with the above protocol achieves
  network stability for any graph $\mathscr{G}\in\mathbb{G}^N$ with
  any size of the network $N$. In other words, the MAS achieves weak
  output synchronization as defined in Lemma \ref{xxxxx2} for all
  graphs of arbitrary size.
\end{problem}

\begin{problem}\label{prob_xcol}
  Consider a MAS \eqref{system} with associated network communication
  \eqref{zeta}, \eqref{zetau}. The \textbf{scale-free collaborative weak output
    synchronization problem} is to find, if possible, a nonlinear
  collaborative protocol of the form \eqref{out_dyncol} using only
  knowledge of agent models, i.e., $(A, B,C)$, such that the MAS with
  the above protocol achieves weak output synchronization for all
  graph $\mathscr{G}\in\mathbb{G}^N$ and with any size of the network
  $N$. In other words, the MAS achieves collaborative weak
  synchronization as defined in Lemma \ref{xxxxx2} for all graphs
  of arbitrary size.
\end{problem}

\section{Scale-free non-collaborative protocol design for MAS}

Next, we design an adaptive non-collaborative protocol to achieve the
objectives of Problem \ref{prob_x}. We make the following assumption. 
\begin{assumption}\label{ass}\mbox{}
  \begin{enumerate}
  \item\label{1.1} $(A,B)$ is stabilizable and $(C,A)$ is detectable.
  \item\label{1.2} $(A,B,C)$ has uniform rank with the order of the
    infinite zeros equal to $1$.
  \item\label{1.4} $(A,B,C)$ is left-invertible.
  \item\label{1.5}  $(A,B,C)$ is minimum-phase.
  \end{enumerate}
\end{assumption}

\begin{remark}
  Assumption \ref{ass}.\ref{1.1} is an obvious necessary condition.

  As argued before, Assumptions \ref{ass}.\ref{1.2} and
  \ref{ass}.\ref{1.5} are quite strong. However, it has been shown in
  \cite{stoorvogel-saberi-liu} that these conditions are very close to
  being necessary in the case of linear protocols. In that paper it is
  also shown that the poles of the system have to be in the closed
  left hald plane but that conditions is removed in this paper by
  extending our class of protocols to include nonlinear
  compensators. We will see later that, using collaborative protocols,
  these assumptions can be relaxed quite significantly.

  Finally Assumptions \ref{ass}.\ref{1.4} can be removed since, when
  $(A, B, C)$ is not left-invertible, we can design a pre-compensator
  to make the system left-invertible. For details we refer to
  \cite{liu-saberi-stoorvogel-tac-2023}.
\end{remark}

\begin{tcolorbox}[breakable,colback=white]
  \noindent\textbf{Protocol 1: Scale-free non-collaborative protocol}
  
  We choose invertible matrices $S$ and $T$ such that
  \[
    \tilde{B}=SB = \begin{pmatrix} 0 \\ B_2 \end{pmatrix},\qquad
    \tilde{C}=TCS^{-1} = \begin{pmatrix} C_1 & 0 \\ 0 & I \end{pmatrix} 
  \]
  with $B_2$ invertible which is possible since $B$ and $CB$ are both
  left-invertible. 
  Define:
  \[
    \begin{pmatrix}
      x_{1i}\\
      x_{2i}
    \end{pmatrix}=Sx_i,\qquad
    \begin{pmatrix}
      y_{1i}\\
      y_{2i}
    \end{pmatrix}=Ty_i,\qquad
    \begin{pmatrix}
      \zeta_{1i}\\
      \zeta_{2i}
    \end{pmatrix}=T\zeta_i,
  \]
  such that the dynamics of $x_{1i}$ and $x_{2i}$ are given
  by
  \begin{equation}\label{eq6}
    \begin{system}{cl}
      \dot{x}_{1i} &= A_{11}x_{1i}+A_{12}x_{2i},\\
      \dot{x}_{2i} &= A_{21}x_{1i}+A_{22}x_{2i}+B_2u_i,\\
      Ty_i&=\begin{pmatrix}
        y_{1i}\\y_{2i}
      \end{pmatrix}=\begin{pmatrix}
        C_1x_{1i}\\ x_{2i}
      \end{pmatrix},
    \end{system}
  \end{equation}
  Since the system is minimum-phase it can be easily verified that we
  must have that $(C_1,A_{11})$ is detectable. Choose $H_1$ such that
  $A_{11}+H_1C_1$ is asymptotically stable
  
  We denote:
  \[
    \tilde{A}=\begin{pmatrix} A_{11} & A_{12} \\ A_{21} &
      A_{22} \end{pmatrix}.
  \]
  Since $(\tilde{A},\tilde{B})$ is stabilizable, there exists a
  matrix $P>0$ satisfying the following algebraic Riccati equation
  \begin{equation}\label{eq-Riccati}
    \tilde{A}\T P + P\tilde{A} -P\tilde{B}\tilde{B}\T P + I =0.
  \end{equation}
  We design the
  following adaptive non-collaborative protocol
  \begin{equation}\label{protocol}
    \begin{system*}{ccl}
      \dot{\hat{\xi}}_{1i} &=& A_{11} \hat{\xi}_{1i} + A_{12} \zeta_{2i}
      + H_1 (C_1\hat{\xi}_{1i} - \zeta_{1i}) \\
      \hat{\xi}_i &=& \begin{pmatrix} \hat{\xi}_{1i} \\ \zeta_{2i} \end{pmatrix} \\
      \dot{\rho}_i &=& \hat{\xi}_i\T
      P\tilde{B}\tilde{B}\T P \hat{\xi}_i, \\
      u_i &=& -\rho_i \tilde{B}\T P \hat{\xi}_i.
    \end{system*}
  \end{equation}
\end{tcolorbox}

\begin{remark}
  In case of full-state coupling (i.e. $C=I$), we have
  \[
    \xi_{1i}=\zeta_{1i},\qquad \text{i.e. we can set}\qquad
    \hat{\xi}_i=\begin{pmatrix}
      \zeta_{1i}\\
      \zeta_{2i}
    \end{pmatrix}=\sum_{j=1}^{N}\ell_{ij}x_{j}
  \]
  which means that we do not need to estimate $\hat{\xi}_{1i}$. With
  $C=I$, Assumptions \ref{ass}, reduced to the coditions that $(A,B)$
  must be stabilizable. Meanwhile, the protocol is designed as
  \[
    \begin{system}{ccl}
      \dot{\rho}_i &=& \zeta_i\T
      P{B}{B}\T P \zeta_i, \\
      u_i &=& -\rho_i {B}\T P \zeta_i.
    \end{system}
  \]
  For details we refer to \cite{stoorvogel-saberi-liu-masoumi}.
\end{remark}

Next, we have the following theorem.
	
\begin{theorem}\label{theorem}
  Consider a MAS \eqref{system}, satisfying assumption \ref{ass}, with
  associated network communication \eqref{zeta}. Then, the
  \textbf{scale-free, non-collaborative weak output synchronization problem} as
  formulated in Problem \ref{prob_x} is solvable. In particular,
  non-collaborative protocol \eqref{protocol} achieves weak output
  synchronization, for an arbitrary number of agents $N$ and for any
  graph $\mathscr{G}\in\mathbb{G}^N$.
\end{theorem}
	
The proof of the above theorem relies on two lemmas which are
presented below. Let us first note the following.  Consider the
non-collaborative protocol \eqref{protocol}. Define
\[
  \xi_i=\sum_{i=1}^N \ell_{ij} Sx_j,\qquad
  \xi_{1i}=\sum_{i=1}^N \ell_{ij} x_{1j},\qquad
  \xi_{2i}=\sum_{i=1}^N \ell_{ij} x_{2j}
\]
Note that we have that $\tilde{C}\xi_i=\zeta_i$,\
$C_1\xi_{1i}=\zeta_{1i}$ and $\xi_{2i}=\zeta_{2i}$.  
We obtain:
\[
  \dot{e}_{1i} = (A_{11}+H_1C_1)e_{1i}
\]
where $e_{1i}= \hat{\xi}_{1i}-\xi_{1i}$. Clearly $e_{1i}$ converges to zero
exponentially and its behavior is independent of our scheduling. We define:
\[
  e_i = \begin{pmatrix} e_{1i} \\ 0 \end{pmatrix},\qquad
  \tilde{E}_2=\begin{pmatrix} H_1C_1 \\ -A_{21} \end{pmatrix}
\]
and
\[
  \hat{\xi}=\begin{pmatrix} \hat{\xi}_1 \\ \vdots \\
    \hat{\xi}_N \end{pmatrix},\qquad
  e=\begin{pmatrix} e_1 \\ \vdots \\
    e_N \end{pmatrix},\qquad
  w=\begin{pmatrix} w_1 \\ \vdots \\
    w_N \end{pmatrix},\qquad 
\]
We obtain:
\begin{equation}\label{closeq}
  \dot{\hat{\xi}} = (I\otimes A)\hat{\xi} - (L\rho\otimes
  \tilde{B}\tilde{B}\T P)\hat{\xi}  +
  (I\otimes \tilde{E}_2) e
\end{equation}

\begin{lemma}\label{lem1}
  Consider a number of agents $N$ and a graph
  $\mathscr{G}\in\mathbb{G}^N$. Consider MAS \eqref{system} with
  associated network communication \eqref{zeta}. Assume Assumption
  \ref{ass} is satisfied.  If all $\rho_i$ remain bounded then
  \begin{equation}\label{probdef}
    \lim_{t\rightarrow \infty}\, \hat{\xi}_i(t)=0,
  \end{equation}
  for all $i=1,\ldots, N$.
\end{lemma}
	
\begin{proof}
  This lemma is basically identical to \cite[Lemma
  3]{stoorvogel-saberi-liu-masoumi}. However, \cite[Equation
  (18)]{stoorvogel-saberi-liu-masoumi} is replaced by \eqref{closeq}
  and the only difference is an exponentially decaying term $e$. The
  proof then follows along the same line as in
  \cite{stoorvogel-saberi-liu-masoumi}.
\end{proof}
	
In the first lemma we showed that if all the adaptive parameters
remain bounded then we obtain our desired result. The next lemma
establishes that the adaptive parameters are actually bounded and we
can use Lemma \ref{lem1} to establish synchronization.
	
\begin{lemma}\label{lem2}
  Consider MAS \eqref{system} with associated network communication
  \eqref{zeta} and the non-collaborative protocol \eqref{protocol}.
  Assume Assumption \ref{ass} is satisfied. Additionally, assume that
  either the $\rho_i$ associated to agents belonging to the basic
  bicomponents are bounded or the graph is strongly connected. In that
  case, all $\rho_i$ remain bounded.
\end{lemma}
	
\begin{proof}
  This lemma is basically identical to \cite[Lemma
  4]{stoorvogel-saberi-liu-masoumi} where the extra exponentially
  decaying term $e$ in \eqref{closeq} requires almost no modification
  of the proof.
\end{proof}\medskip
	
\begin{proof}[Proof of Theorem \ref{theorem}]
  We can again basically follow the same arguments as in the proof of
  \cite[Theorem 1]{stoorvogel-saberi-liu-masoumi}
\end{proof}

\section{Scale-free collaborative protocol design for MAS}
	
Next, we design an adaptive collaborative protocol to achieve the
objectives of Problem \ref{prob_xcol}. We make the following
assumptions:
	
\begin{assumption}\label{asscol}\mbox{}
\item $(A,B)$ is stabilizable and $(C,A)$ is observable.
\end{assumption}

\begin{remark}
  Compared to non-collaborative protocols all restrictive assumptions
  have been removed. Stabilizability and detectability are clearly
  necessary.
\end{remark}
	
\begin{remark}
  In case of full-state coupling (i.e. $C=I$), it is obvious that the
  collaborative protocol design only needs $(A,B)$ is stabilizable.
\end{remark}
		
Next, we describe the collaborative protocol we will be using in this
paper.
	
\begin{tcolorbox}[breakable, colback=white]		
  \noindent\textbf{Protocol 2: Scale-free collaborative protocol}
		
  Choose $Q>0$ such that:
  \[
    AQ+QA\T-QC\T CQ+I = 0
  \]
  We will use the following observer:
  \begin{equation} \label{protocol2.1}
    \dot{\hat{x}}_i=A\hat{x}_i+Bu_i-\rho_i QC\T
    (C\tilde{\zeta}_i-\zeta_i)
  \end{equation}
  with $\zeta_i$ given by \eqref{zeta} and $\tilde{\zeta}_i$ given by \eqref{zetau}
  with $x_{i,c}=\hat{x}_i$. Define:
  \[
    e_i=\hat{x}_i-x_i,\qquad \xi_i = \sum_{j=1}^N \ell_{ij} e_j,\qquad
    \tilde{e}_i=\sum_{j=1}^N \ell_{ij} Ce_j  
  \]
  We have:
  \[
    \tilde{e}_i = C\tilde{\zeta}_i-\zeta_i = C\xi_i.
  \]
  and we choose the following adaptive gain:
  \begin{equation} \label{protocol2.2}
    \dot{\rho}_i = 
    \tilde{e}_i\T \tilde{e}_i.
  \end{equation}
  Finally, we set
  \begin{equation} \label{protocol2.4}
    u_i = F\hat{x}_i
  \end{equation}
  where $A+BF$ is asymptotically stable.
\end{tcolorbox}

We have the following theorem.
	
\begin{theorem}\label{theorem2}
  Consider a MAS \eqref{system}, satisfying assumption \ref{asscol},
  with associated network communication \eqref{zeta}. Then, the
  \textbf{scale-free, collaborative weak output synchronization problem} as
  formulated in Problem \ref{prob_xcol} is solvable. In particular, the
  collaborative protocol given by \eqref{protocol2.1},
  \eqref{protocol2.2} and \eqref{protocol2.4} achieves weak output
  synchronization for an arbitrary number of agents $N$ and for any
  graph $\mathscr{G}\in\mathbb{G}^N$.
\end{theorem}
	
Part of the proof of the above theorem is given in two lemmas to
improve the presentation. These lemmas establish the behavior of the
observer \eqref{protocol2.1} with its adaptive gain given by
\eqref{protocol2.2}
	
From the observer, we obtain:
\begin{equation} \label{syscomp}
  \dot{\xi}_i=A\xi_i-(L_i\rho \otimes QC\T C)\xi 
\end{equation}
where $L_i$ is the $i$'th row of $L$ for $i=1,\ldots, N$ and
\[
  \xi=\begin{pmatrix} \xi_1 \\ \xi_2 \\ \vdots \\ \xi_N 
  \end{pmatrix},\qquad
  \rho= \begin{pmatrix}
    \rho_1  & 0          & \cdots & 0      \\
    0         & \rho_2 & \ddots & \vdots \\
    \vdots    & \ddots     & \ddots   & 0 \\
    0         & \cdots     & 0        & \rho_N
  \end{pmatrix}.
\]
together with the adaptive gain \eqref{protocol2.2}. We would like to
stress the crucial feature that this dynamics is completely
independent of the state feedback \eqref{protocol2.4}. Therefore the
behavior of \eqref{syscomp} can be analyzed as a closed unit
independent of the rest of the dynamics.
	
\begin{lemma}\label{lem1col}
  Consider an arbitrary number of agents $N$ and an arbitrary graph
  $\mathscr{G}\in\mathbb{G}^N$. Consider the MAS \eqref{system} with
  associated network communication \eqref{zeta}. Assume Assumption
  \ref{asscol} is satisfied.  If all $\rho_i$ remain bounded then
  \begin{equation}\label{probdef2}
    \lim_{t\rightarrow \infty} \zeta_i(t)=0,\qquad
    \lim_{t\rightarrow \infty} \tilde{\zeta}_i(t)=0.
  \end{equation}
  for all $i=1,\ldots, N$.
\end{lemma}

\begin{proof}[Proof of Lemma \ref{lem1col}]
  For the system \eqref{syscomp}, we define
  $V_i = \xi_i\T Q^{-1} \xi_i$ and we obtain:
  \[
    \dot{V}_i = \xi_i\T(-\eta Q^{-1}+C\T C)\xi_i - 2\xi_i\T (L_i\rho
    \otimes C\T C)\xi 
  \]
  where $\eta=\| Q \|^{-1}$ and hence:
  \begin{equation}\label{dfg1}
    \dot{V}_i \leq - \tfrac{\eta}{2} V_i + \tilde{e}_i\T \tilde{e}_i - 2\tilde{e}_i\T
    (L_i\rho \otimes I) \tilde{e} 
  \end{equation}
  with $\tilde{e}=(I\otimes C)\xi$. Convergence of $\rho_i$ for
  $i=1,\ldots, N$ implies that $\tilde{e}_i\in L_2$. Equation
  \eqref{dfg1} then implies:
  \begin{equation}\label{asd1}
    \lim_{t\rightarrow\infty} \xi_i(t) =0
  \end{equation}
  Equation \eqref{protocol2.1} with $u_i$ given by \eqref{protocol2.4}
  then implies
  \[
    \dot{\hat{x}}_i=(A+BF)\hat{x}_i-\rho_i QC\T \tilde{e}_i
  \]
  and hence
  \[
    \lim_{t\rightarrow\infty} \hat{x}_i(t) =0
  \]
  since $A+BF$ is asymptotically stable which implies:
  \begin{equation}\label{asd2}
    \lim_{t\rightarrow\infty} \tilde{\zeta}_i(t) =0
  \end{equation}
  Combining \eqref{asd1} and \eqref{asd2} then yields \eqref{probdef2}
  which completes the proof.
\end{proof}

\begin{lemma}\label{lem2col}
  Consider a number of agents $N$ and a graph
  $\mathscr{G}\in\mathbb{G}^N$. Consider the MAS \eqref{system} with
  associated network communication \eqref{zeta}. Assume Assumption
  \ref{asscol} is satisfied. Additionally, assume that either the
  $\rho_i$ associated to agents belonging to the basic bicomponents
  are bounded or the graph is strongly connected. In that case, all
  $\rho_i$ remain bounded.
\end{lemma}
	
\begin{proof}[Proof of Lemma \ref{lem2col}]
  Using the notation of Lemma \ref{lem1col} we obtain \eqref{syscomp}
  for $i=1,\ldots,N$ or equivalently:
  \begin{equation}\label{syscomp2}
    \dot{e} =(I\otimes A)e -(\rho L \otimes QC\T C)e.
  \end{equation}
  where
  \[
    e=\begin{pmatrix} e_1 \\ \vdots \\ e_N \end{pmatrix},\quad
    w=\begin{pmatrix} w_1 \\ \vdots \\ w_N \end{pmatrix},
  \]
  If all $\rho_i$ are bounded the result of the lemma is
  trivial. Assume that $k\leq N$ of the $\rho_i$ are
  unbounded. Without loss of generality, we assume that the $\rho_i$
  are unbounded for $i\leq k$ while the $\rho_i$ are bounded for
  $i>k$.
		
  We first consider the case that $k<N$. We have
  \begin{equation*}
    L = \begin{pmatrix}
      L_{11} & L_{12} \\ L_{21} & L_{22} 
    \end{pmatrix},\quad
    e^k =\begin{pmatrix} e_1 \\ \vdots \\ e_k \end{pmatrix}, \quad
    e^k_c =\begin{pmatrix} e_{k+1} \\ \vdots \\
      e_N \end{pmatrix},\quad
    \tilde{e}^k =\begin{pmatrix} \tilde{e}_{1} \\ \vdots \\
      \tilde{e}_{k} \end{pmatrix},\quad
    \tilde{e}^k_c =\begin{pmatrix} \tilde{e}_{k+1} \\ \vdots \\
      \tilde{e}_{N} \end{pmatrix},
  \end{equation*}
  with $L_{11}\in \R^{k\times k}$. If all the agents associated to
  basic bi-components have a bounded $\rho_i$ this implies that agents
  associated to $i=1,\ldots,k$ are not associated to basic
  bi-components which implies that $L_{11}$ is invertible. On the
  other hand, if the network is strongly connected we always have that
  $L_{11}$ is invertible since $k<N$. Since the $\rho_i$ are bounded
  for $i>k$ we find that $\tilde{e}^k_c\in L_2$. We define
  \[
    \hat{e}^k= e^k+(L_{11}^{-1}L_{12}\otimes I)
    e^k_c,\qquad 
    \tilde{e}^k=(L_{11}\otimes C)\hat{e}^k
  \] 
  Using \eqref{syscomp2}, we then obtain
  \begin{equation}\label{barxt2xxx}
    \dot{\hat{e}}^k = (I\otimes A)\hat{e}^k
    -[\rho^{k} L_{11} \otimes QC\T C] \hat{e}^k
    -\left[ L_{11}^{-1}L_{12}\rho^k_c
      \otimes QC\T \right]e^k_c
  \end{equation}
  where we used that
  \[
    \rho^k=\begin{pmatrix}
      \rho_1 & 0      & \cdots & 0 \\
      0      & \rho_2 & \ddots & \vdots \\
      \vdots & \ddots & \ddots & 0 \\
      0      & \cdots & 0      & \rho_k
    \end{pmatrix},\quad
    \rho^k_c=\begin{pmatrix}
      \rho_{k+1} & 0      & \cdots & 0 \\
      0      & \rho_{k+2} & \ddots & \vdots \\
      \vdots & \ddots & \ddots & 0 \\
      0      & \cdots & 0      & \rho_N
    \end{pmatrix}.
  \]
  Define
  \[
    \hat{v}^k =-(L_{11}^{-1}L_{12}\rho^k_c \otimes QC\T)e^k_c , 
  \]
  then the boundedness of $\rho^k_c$ implies that there exists $K$
  such that
  \begin{equation}\label{K3K4}
    \| \hat{v}^k \|_2 < K.
  \end{equation}
  For $k<N$, we set
  \begin{equation}\label{tildeVk}
    V_{k}= (\hat{e}^k)\T (\rho^{-k} H^k \otimes
    Q^{-1}) \hat{e}^k, 
  \end{equation}
  with $\rho^{-k}=(\rho^k)^{-1}$
  while
  \begin{equation}\label{HN}
    H^k=\begin{pmatrix}
      h_1 & 0      & \cdots & 0 \\
      0      & h_2 & \ddots & \vdots \\
      \vdots & \ddots & \ddots & 0 \\
      0      & \cdots & 0      & h_k
    \end{pmatrix}.
  \end{equation}
  where, using \cite[Theorem 4.25]{qu-book-2009}, we choose
  $h_1,\ldots,h_k>0$ such that $H^k L_{11} +L_{11}\T H^k > 0$. It is
  easily seen that this implies that there exists a $\gamma$ such that
  \begin{equation}\label{HkL11}
    H^kL_{11}+L_{11}\T H^k > 2\gamma L_{11}\T L_{11}.
  \end{equation}
  Choose $T$ such that
  \[
    \rho^{-k} H^k  < \gamma L_{11}\T L_{11},
  \]
  for $t>T$ which is possible since we have $\rho_i\rightarrow \infty$
  for $i=1,\ldots, k$.  Note that given \eqref{K3K4} there exists some
  fixed $\beta$ such that
  \begin{equation}\label{lastone4xx}
    \| \rho^k \check{v}^k \|  \leq \beta,
  \end{equation}
  with
  \[
    \check{v}^k = \left[ \tfrac{2}{\eta} \rho^{-k}H^k \otimes
      Q^{-1}\right]^{1/2} \hat{v}^k.
  \]
  where $\eta=\| Q \|^{-1}$. We get
  \begin{equation}\label{lastone2xx}
    \dot{V}_k \leq - \tfrac{\eta}{2} V_k 
    - \gamma (\hat{e}^k)\T \left[ L_{11}\T L_{11} \otimes C\T C \right] \hat{e}^k 
    + (\check{v}^k)\T\check{v}^k,
  \end{equation}
  for $t>T$. We have:
  \begin{equation}\label{rfgt}
    (\hat{e}^k)\T \left[ L_{11}\T L_{11} \otimes C\T C \right]
    \hat{e}^k =  \sum_{i=1}^k \dot{\rho}_i, 
  \end{equation}
  since $(L_{11}\otimes C)\hat{e}^k=\tilde{e}^k$. Hence \eqref{lastone2xx} implies
  \begin{equation}\label{ttg1}
    \dot{V}_k \leq -\tfrac{\eta}{2} V_k -\gamma
    \sum_{i=1}^k \dot{\rho}_i  +
    (\check{v}^k)\T\check{v}^k.  
  \end{equation}
  Since the $\rho_i$ are unbounded while $\check{v}^k\in L_2$ this
  yields a contradiction since $V_k\geq 0$.
  
  Next, we consider the case that all $\rho_i$ are unbounded. In this
  case, we assumed the graph is strongly connected and hence by Lemma
  \ref{2.8} presented in the appendix there exists
  $\alpha_1,\ldots,\alpha_N>0$ such that \eqref{Hlyap} is satisfied
  with $H^N$ given by \eqref{HN} for $k=N$.  We define
  \begin{equation}\label{VN}
    V_N= e\T \left[ Q_{\rho} \otimes Q^{-1} 
    \right] e,
  \end{equation}
  where
  \begin{equation}\label{Qrho}
    Q_{\rho} = \rho^{-1} \left( H^N\rho - \mu_N
      \textbf{h}_N\textbf{h}_N\T \right) \rho^{-1}
  \end{equation}
  while
  \[
    \mu_N=\frac{1}{\sum_{i=1}^N h_i\rho_i^{-1}},\qquad \textbf{h}_N
    = \begin{pmatrix} h_1 \\ \vdots \\ h_N \end{pmatrix}.
  \]
  From Lemma \ref{2.9} in the appendix, we know that $Q_{\rho}$ is
  decreasing in $\rho_i$ for $i=1,\ldots N$.  Note that
  $Q_\rho \rho L=H^N L$.  We get from \eqref{syscomp2} that
  \begin{equation}\label{eqref}
    \dot{V}_N \leq e\T \left[ Q_\rho \otimes (-\eta Q^{-1}+C\T C) \right] e
    -e\T \left[ (H^NL+L\T H^N) \otimes C\T C \right] e 
  \end{equation}
  It is easily verified that $Q_{\rho}\textbf{1}=0$. Moreover
  $\Ker L = \Span \{ \textbf{1} \}$ since the network is strongly
  connected. Therefore
  \[
    \ker Q_{\rho} \subset \Ker L\T L.
  \]
  Together with the fact that $\rho_j\rightarrow \infty$ for
  $j=1,\ldots, N$ and therefore $Q_{\rho}\rightarrow 0$ this implies
  that there exists $T$ such that
  \begin{equation}\label{Qrhobound}
    Q_{\rho} < \gamma L\T L,
  \end{equation}
  is satisfied for $t>T$.
		
  The above together with \eqref{Hlyap} yields for $t>T$ that
  \begin{equation*}
    \dot{V}_N \leq -\eta V_N
    - \gamma e\T \left[ L\T L \otimes C\T C \right] e \\
  \end{equation*}
  We get
  \begin{equation}\label{lastone2xxN}
    \dot{V}_N \leq - \tfrac{\eta}{2} V_N
    - \gamma e\T \left[ L\T L \otimes C\T C \right] e 
  \end{equation}
  for $t>T$. Moreover,
  \[
    e\T \left[ L\T L \otimes C\T C \right] e = \sum_{i=1}^N
    \dot{\rho}_i.
  \]
  The above implies
  \begin{equation}\label{ttg1N}
    \dot{V}_N \leq - \tfrac{\eta}{2} V_N -\gamma
    \sum_{i=1}^N \dot{\rho}_i. 
  \end{equation} 
  We find that since the $\rho_i$ are unbounded yields a contradiction
  since $V_k\geq 0$.
  
  Therefore both for $k<N$ and $k=N$ we obtain a contradiction with
  the assumption that some of the $\rho_i$ are unbounded. This
  completes the proof of the lemma.
\end{proof}\medskip
	
\begin{proof}[Proof of Theorem \ref{theorem2}]
  The Laplacian matrix of the system in general has the form
  \eqref{Lstruc}. We note that if we look at the dynamics of the
  agents belonging to one of the basic bicomponents then these
  dynamics are not influenced by the other agents and hence can be
  analyzed independent of the rest of network. The network within one
  of the basic bicomponents is strongly connected and we can apply
  Lemma \ref{lem2col} to guarantee that the $\rho_i$ associated to a
  basic bicomponents are all bounded.
		
  Next, we look at the full network again. We have already established
  that the $\rho_i$ associated to all basic bicomponents are all
  bounded. Then we can again apply Lemma \ref{lem2col} to conclude
  that the other $\rho_i$ not associated to basic bicomponents are
  also bounded.
		
  After having established that all the $\rho_i$ are all bounded, we
  can then apply Lemma \ref{lem1col} to conclude that \eqref{probdef2}
  is satisfied which completes the proof of the theorem.
\end{proof}
	
\section{Numerical examples}

In this section, we provide two examples to achieve the weak output
synchronization by the adaptive non-collaborative and collaborative
protocols, respectively.

\subsection{Networks without a spanning tree}
	
\begin{figure}[t]
  \includegraphics[width=6cm]{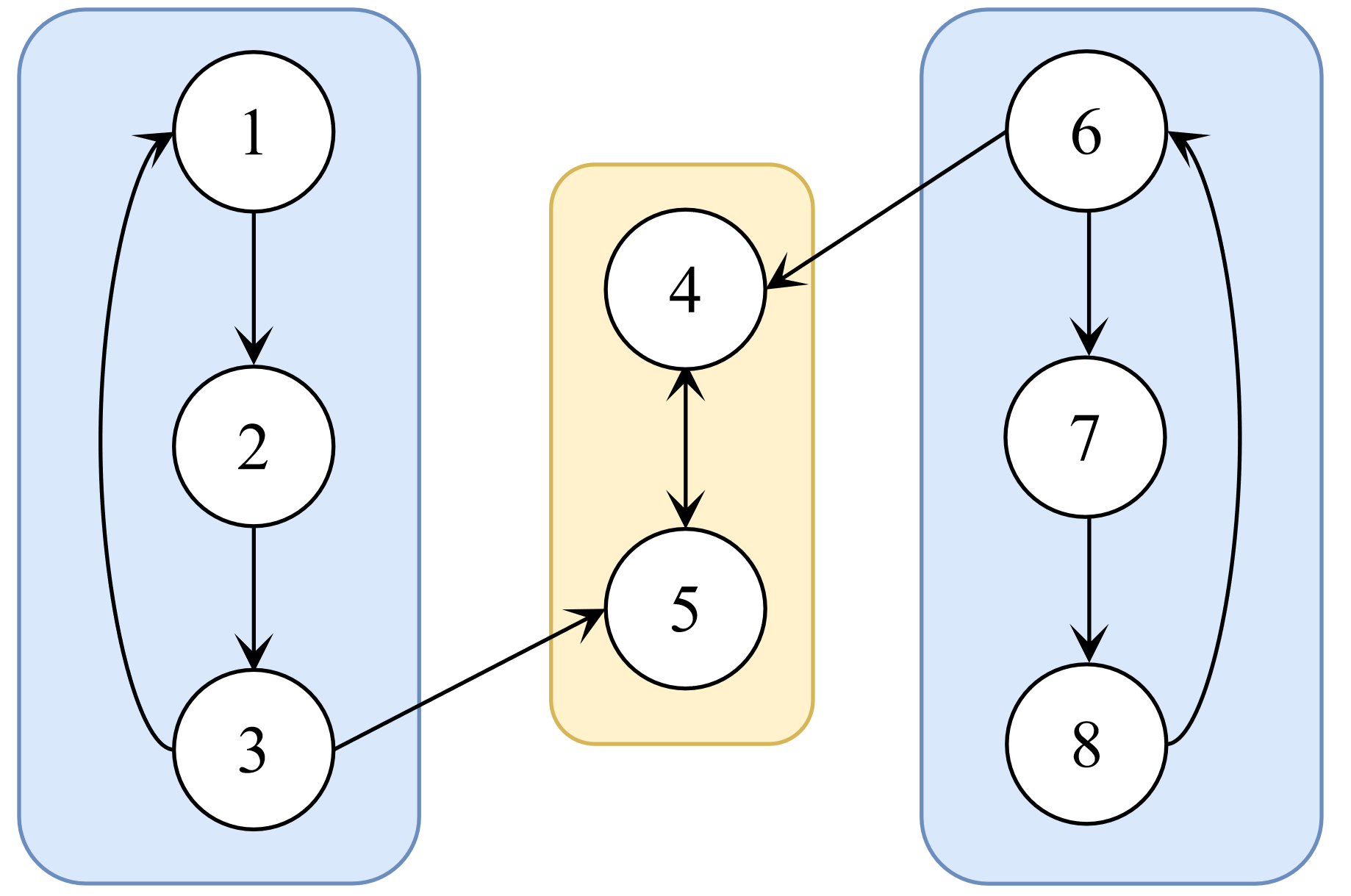} \centering
  \caption{The 8-node communication network without spanning
    tree.}\label{f3}
\end{figure}

We consider two different networks in this section.  Figure \vref{f3}
depicts a small network with two basic bicomponents (indicated in
color blue): Cluster 1 containing 3 nodes and Cluster 2 containing 3
nodes. Moreover, we have one non-basic bicomponent: Cluster 3
containing 2 nodes, which are indicated in yellow. Clearly the network
has no spanning tree.

\begin{figure}[t]
  \includegraphics[width=6cm]{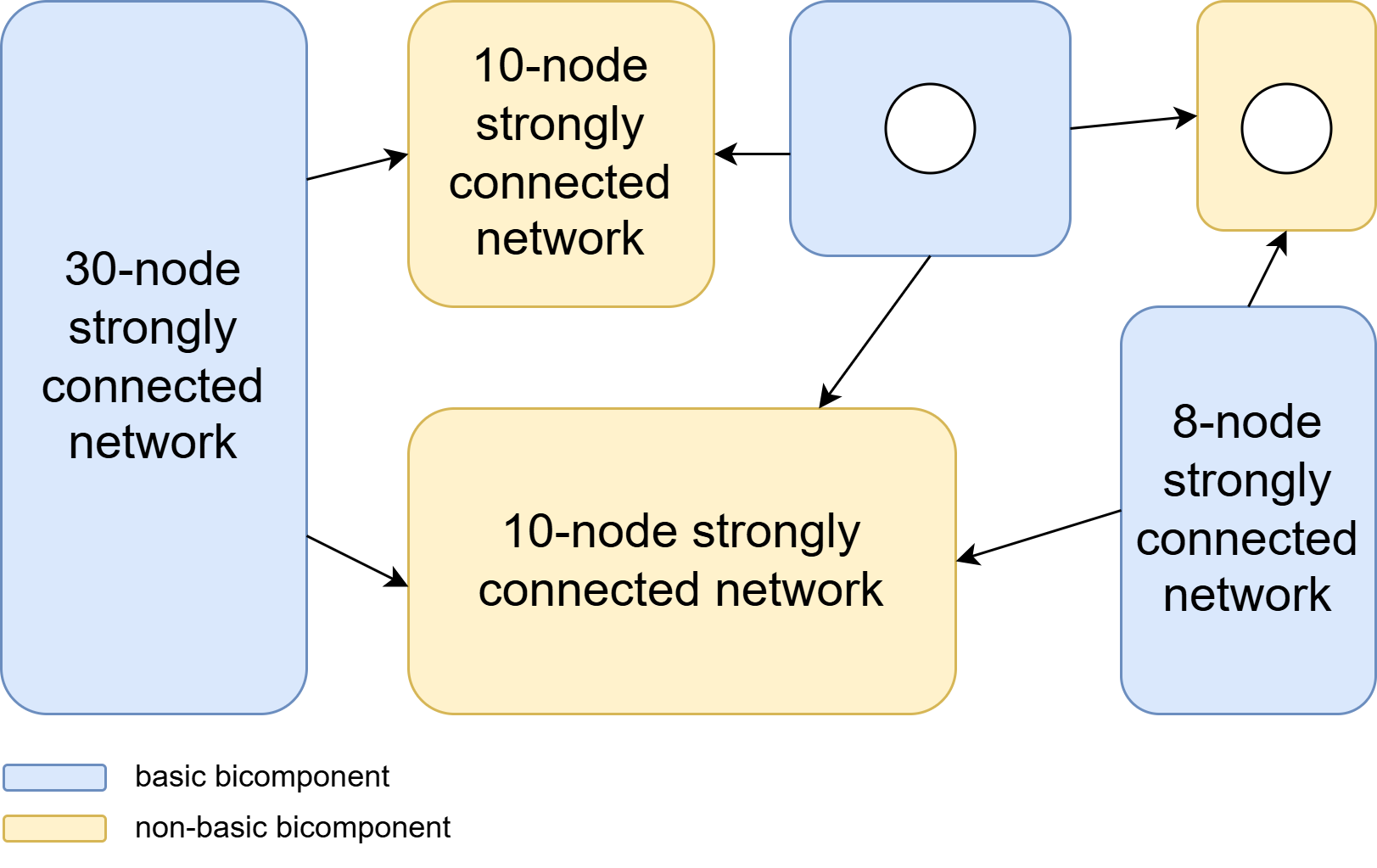} \centering
  \caption{The 60-nodes communication network without spanning
    tree.}\label{f4}
\end{figure}

The second network depicted in Figure \vref{f4} is a larger network
with three basic bicomponents (indicated in blue): Cluster 1
containing 30 nodes, Cluster 2 containing 8 nodes, and Cluster 3
containing $1$ node. Moreover, we have three non-basic bicomponents:
Cluster 4 containing 10 nodes, Cluster 5 containing 10 nodes, and
Cluster 6 containing 1 node, which are indicated in yellow.
	
\subsection{Non-collaborative protocol}
	
We consider again MAS \eqref{system} consisting of neutrally stable agents given by:
\begin{equation}\label{IE1.1}
  A=\begin{pmatrix}
    0&1&1\\
    -1&0&1\\
    0&0&0
  \end{pmatrix}, B=\begin{pmatrix}
    0\\0\\1
  \end{pmatrix}, C=\begin{pmatrix}
    1&0&0\\
    0&0&1
  \end{pmatrix}.
\end{equation}
For this agent, we can set $S=I_{3}, T=I_{2}$. Then we can design a adaptive
non-collaborative protocol \eqref{protocol} with
\begin{equation}\label{IE1.2}
  H_{1}=\begin{pmatrix}
    -3\\-1
  \end{pmatrix}, P=\begin{pmatrix}
    2.5958  &  0.2321  &  0.7321\\
    0.2321  &  1.7100  &  1.2100\\
    0.7321  &  1.2100  &  2.2100
  \end{pmatrix}.
\end{equation}
by solving algebraic Riccati equation \eqref{eq-Riccati}.

\begin{figure}[ht]
  \includegraphics[width=8cm]{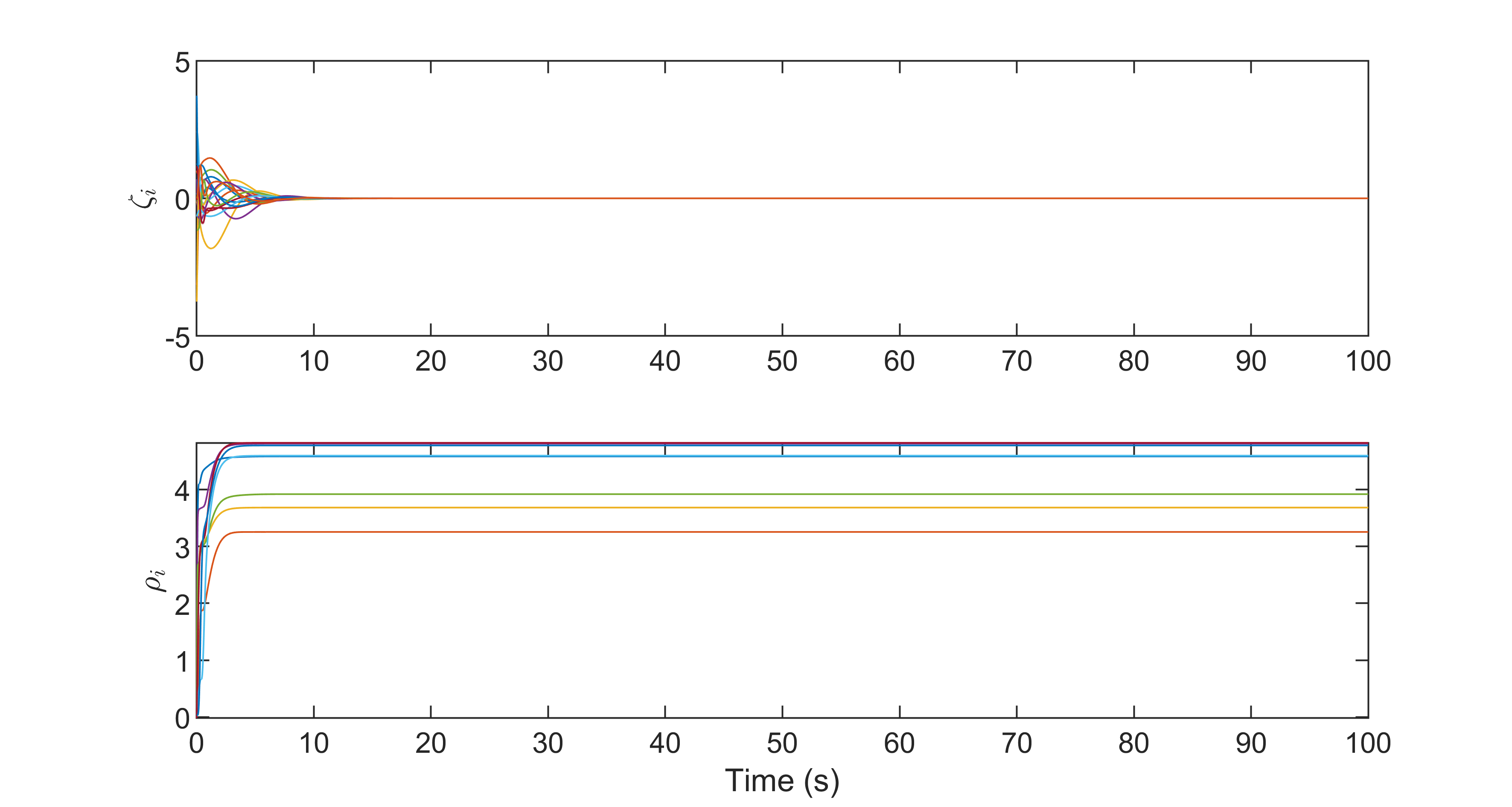}
  \centering
  \caption{The network information $\zeta_{i}$ and the adaptive
    parameters $\rho_{i}$ for 8-node network without a spanning tree
    via adaptive non-collaborative protocol.}
  \label{8nonc_wsyn_zeta_rho1}
\end{figure}

\begin{figure}[ht]
  \includegraphics[width=8cm]{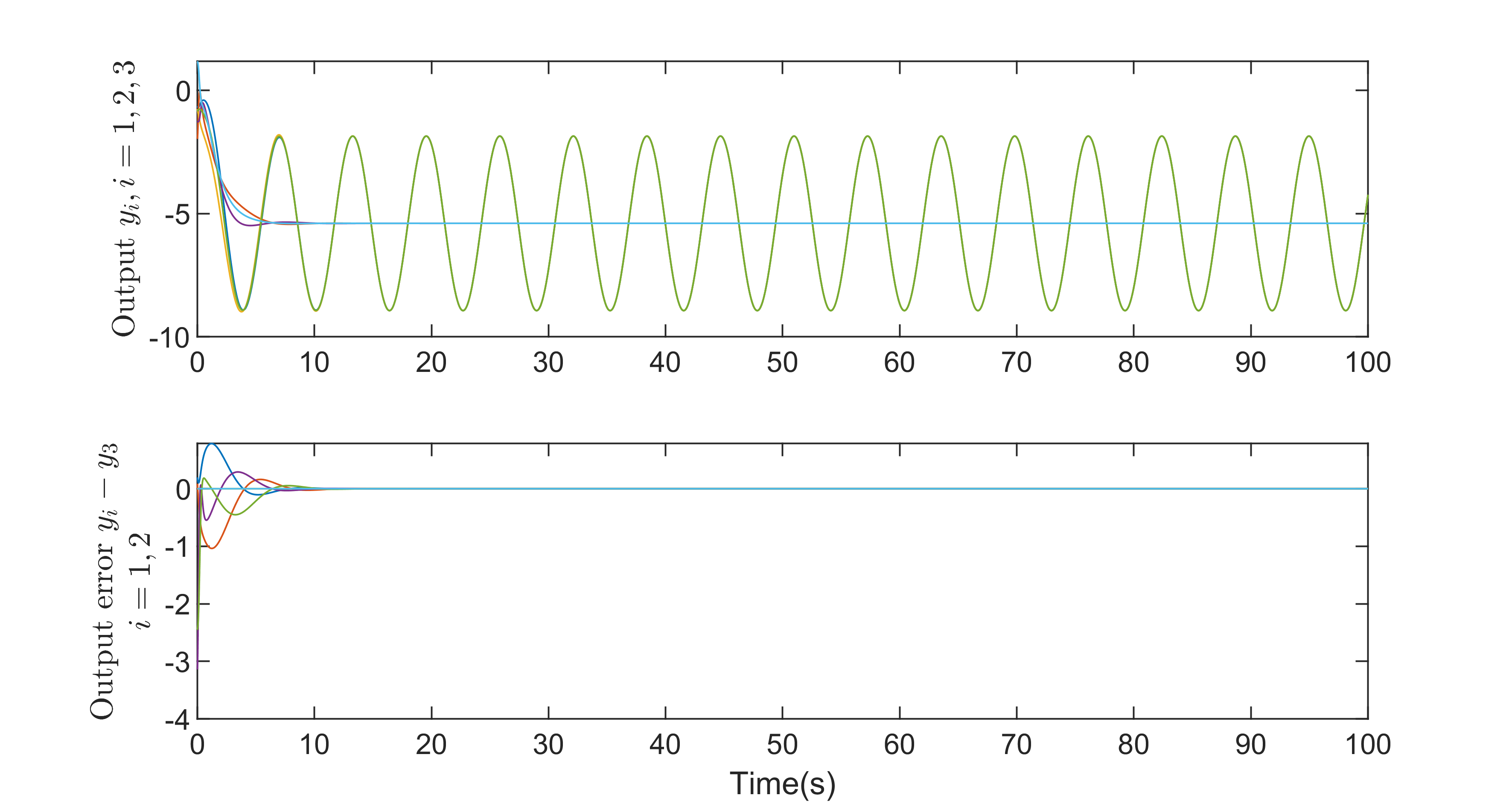}
  \centering
  \caption{The output synchronization and error output results of
    $y_i, i=1,2,3$ for 8-node network without a spanning tree via
    adaptive non-collaborative protocol.}
  \label{8nonc_wsyn_output123_error}
\end{figure}

\begin{figure}[ht]
  \includegraphics[width=8cm]{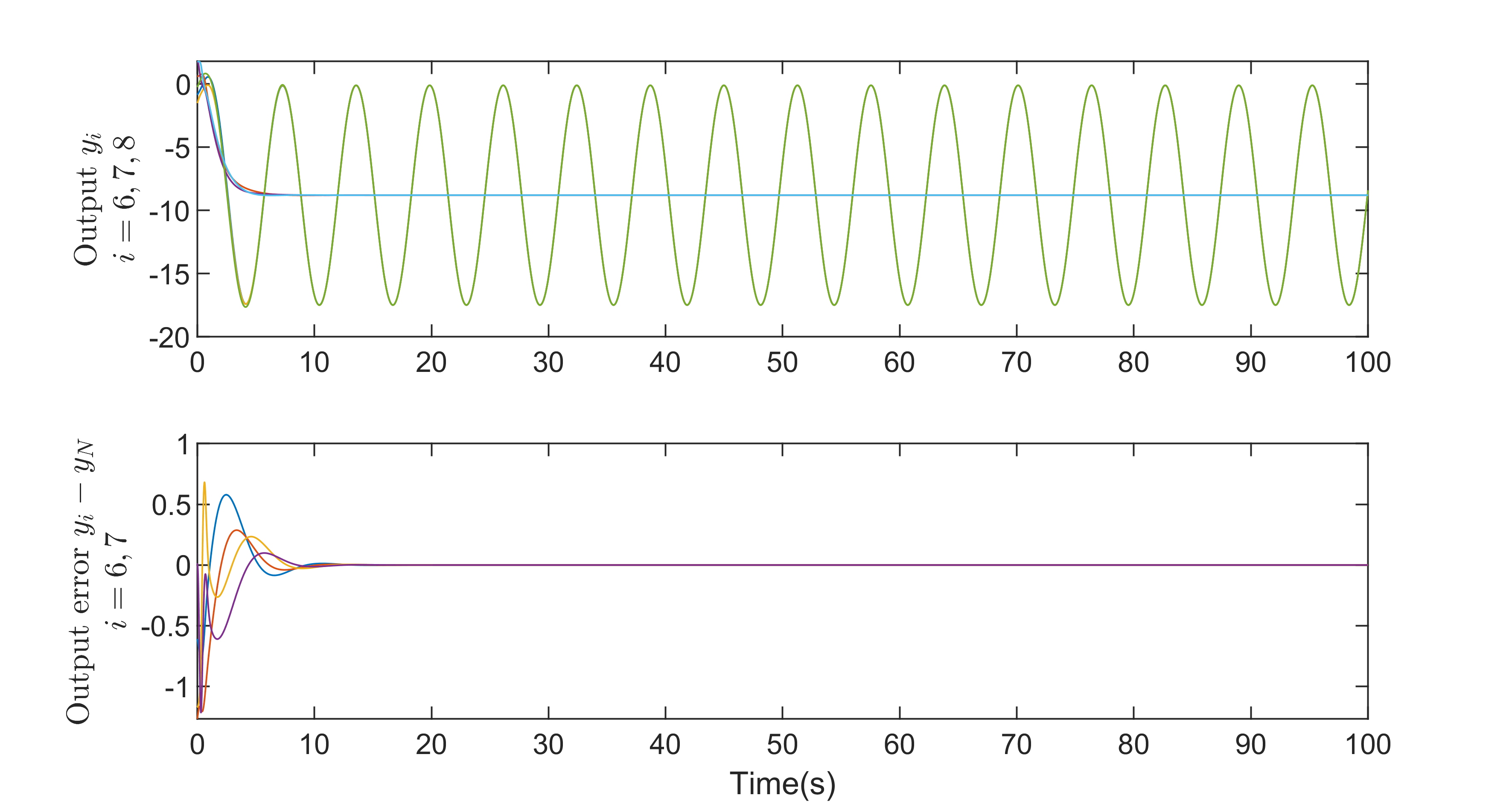}
  \centering
  \caption{The output synchronization and error output results of
    $y_i, i=6,7,8$ for 8-node network without a spanning tree via
    adaptive non-collaborative protocol.}
  \label{8nonc_wsyn_output678_error}
\end{figure}

By using the above protocol, we obtain weak output
synchronization for the network given by Figures \ref{f3} and
\ref{f4}. For the network shown in Figure \ref{f3}, $\zeta_{i}$ and
$\rho_{i}$ are given in Figure \ref{8nonc_wsyn_zeta_rho1}. We note
that we indeed achieve output synchronization for Clusters 1 and 2,
which describe the two basic bicomponents, as shown in Figures
\ref{8nonc_wsyn_output123_error} and \vref{8nonc_wsyn_output678_error}.

\begin{figure}[htp]
  \includegraphics[width=8cm]{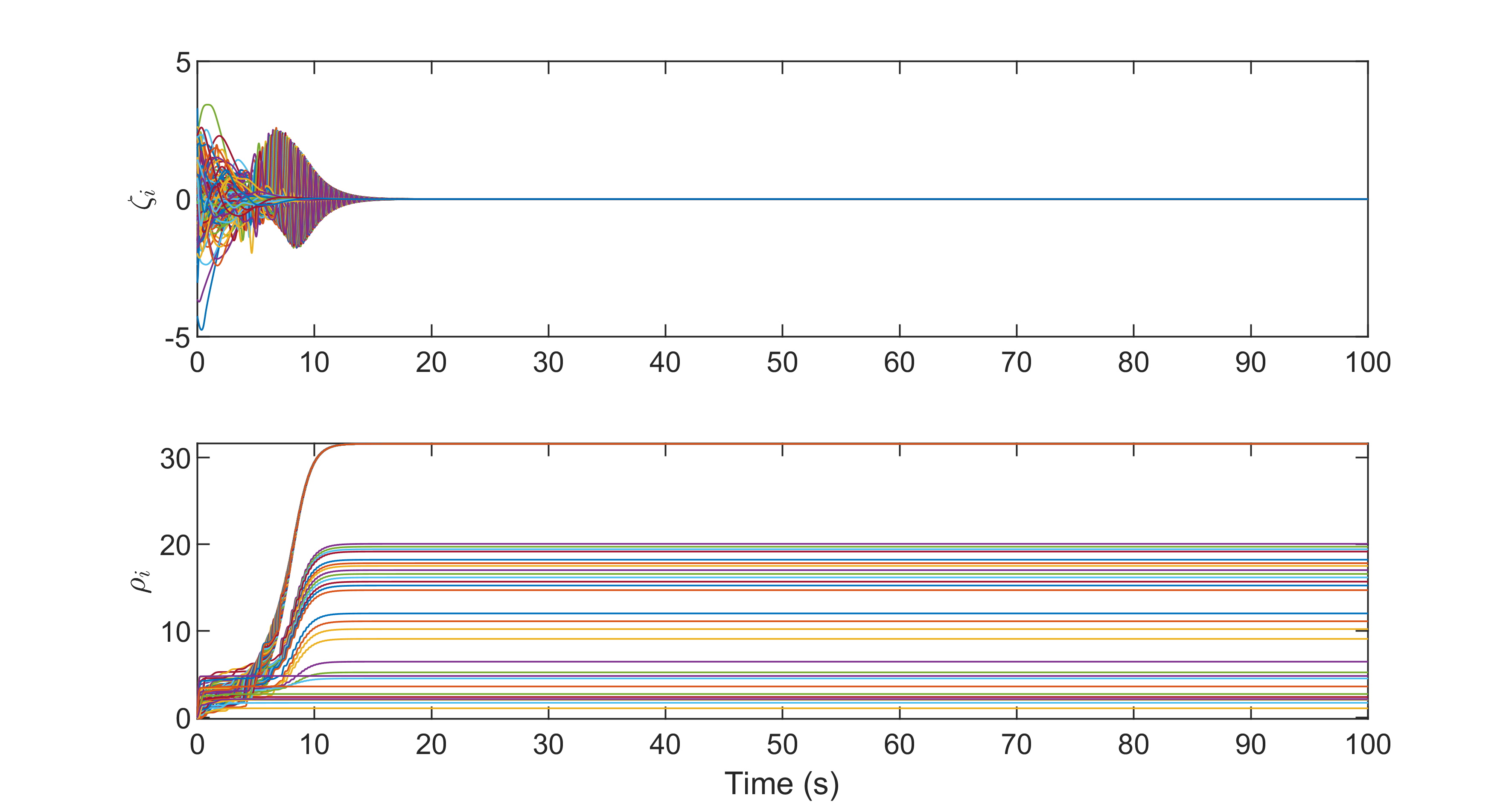} \centering
  \caption{The network information $\zeta_{i}$ and the adaptive
    parameters $\rho_{i}$ for 60-node network without a spanning tree
    via adaptive non-collaborative protocol.}
  \label{60nonc_wsyn_zeta_rho1}
\end{figure}

\begin{figure}[htp]
  \includegraphics[width=8cm]{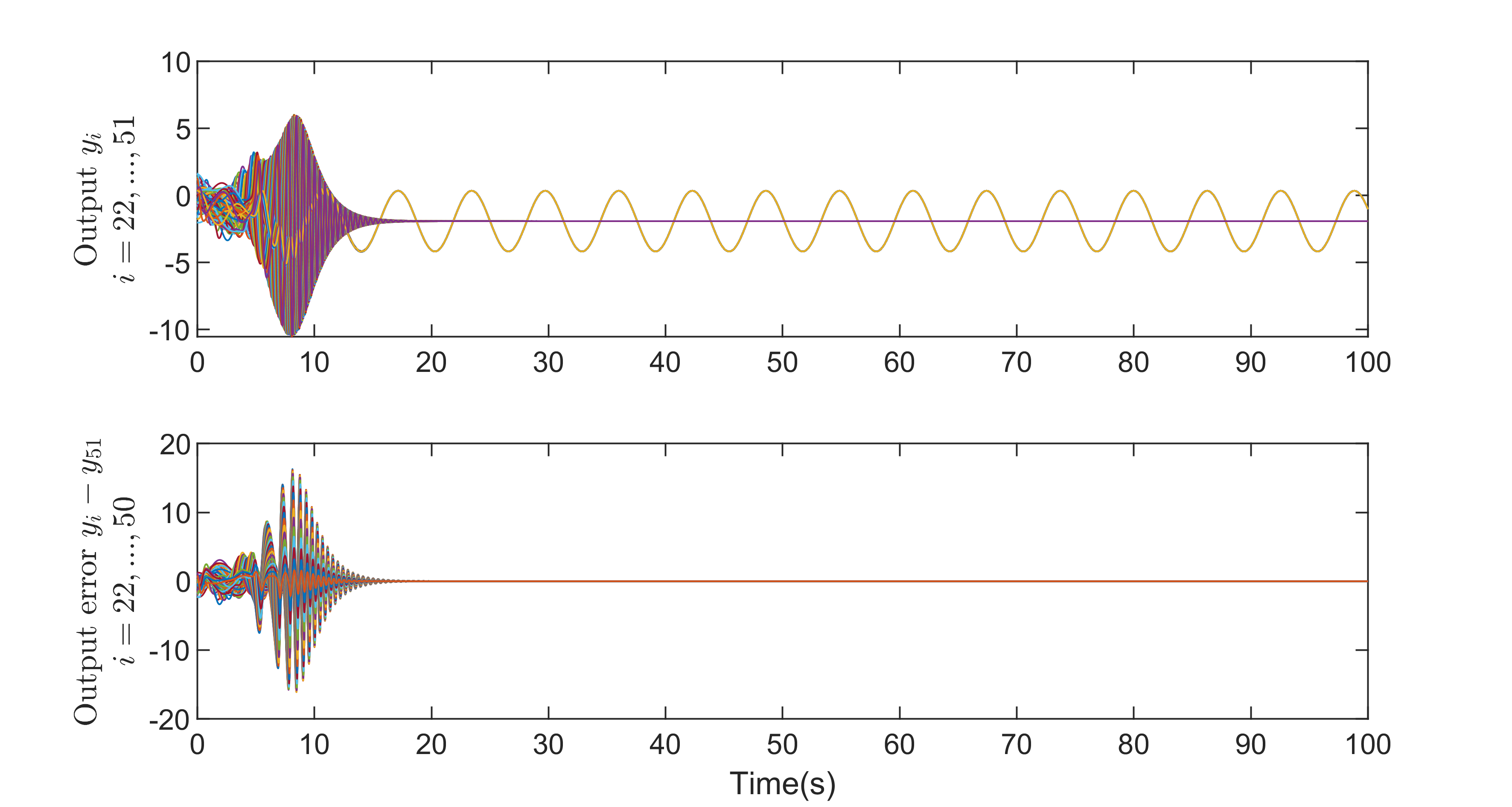}
  \centering
  \caption{The output synchronization and error state results of
    $y_i, i=22,\ldots,51$ for 60-node network without a spanning tree
    via adaptive non-collaborative protocol.}
  \label{60nonc_wsyn_output22-51_error}
\end{figure}

\begin{figure}[htp]
  \includegraphics[width=8cm]{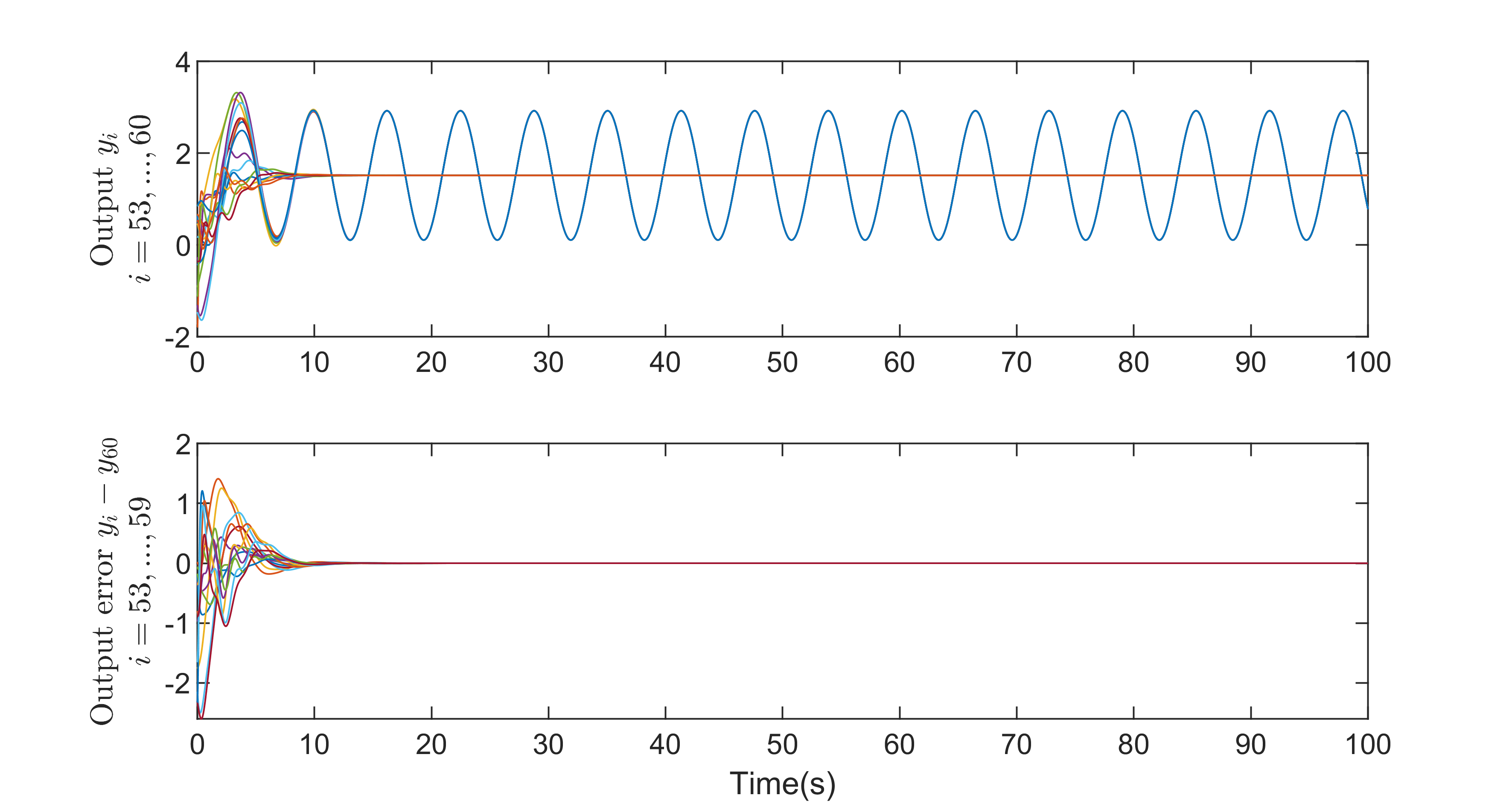}
  \centering
  \caption{The output synchronization and error state results of
    $y_i, i=53,\ldots,60$ for 60-node network without a spanning tree
    via adaptive non-collaborative protocol.}
  \label{60nonc_wsyn_output53-60_error}
\end{figure}

For the network shown in Figure \ref{f4}, $\zeta_{i}$ and the adaptive
parameters $\rho_{i}$ are given in Figure \vref{60nonc_wsyn_zeta_rho1}.
The synchronization output trajectories and errors for Clusters 1 and
2 are shown in Figures \ref{60nonc_wsyn_output22-51_error} and
\vref{60nonc_wsyn_output53-60_error} and we see that we indeed achieve
output synchronization. Since Cluster 3 only has one node, we do not
need to provide its trajectory.
	
\FloatBarrier 

\subsection{Collaborative protocol}
	
For collaborative protocol case, we consider a MAS \eqref{system} with
the following parameters:
\begin{equation}\label{IE2.1}
  A=\begin{pmatrix}
    0&1&0\\
    0&0&1\\
    0&0&0
  \end{pmatrix}, B=\begin{pmatrix}
    0\\0\\1
  \end{pmatrix}, C=\begin{pmatrix}
    1&0&0
  \end{pmatrix}.
\end{equation}

Meanwhile, we design the adaptive collaborative protocol
\eqref{protocol2.1}-\eqref{protocol2.4} with
\begin{equation}\label{IE2.2}
  F=\begin{pmatrix}
    -6\\-11\\-6
  \end{pmatrix}, Q=\begin{pmatrix}
    2.4142  &  2.4142  &  1.0000\\
    2.4142  &  4.8284  &  2.4142\\
    1.0000  &  2.4142  &  2.4142
  \end{pmatrix}.
\end{equation}
For the adaptive collaborative protocol, we still consider the
previous 8-node and 60-node networks given by Figures \ref{f3} and
\ref{f4}.
	
\begin{figure}[ht]
  \includegraphics[width=8cm]{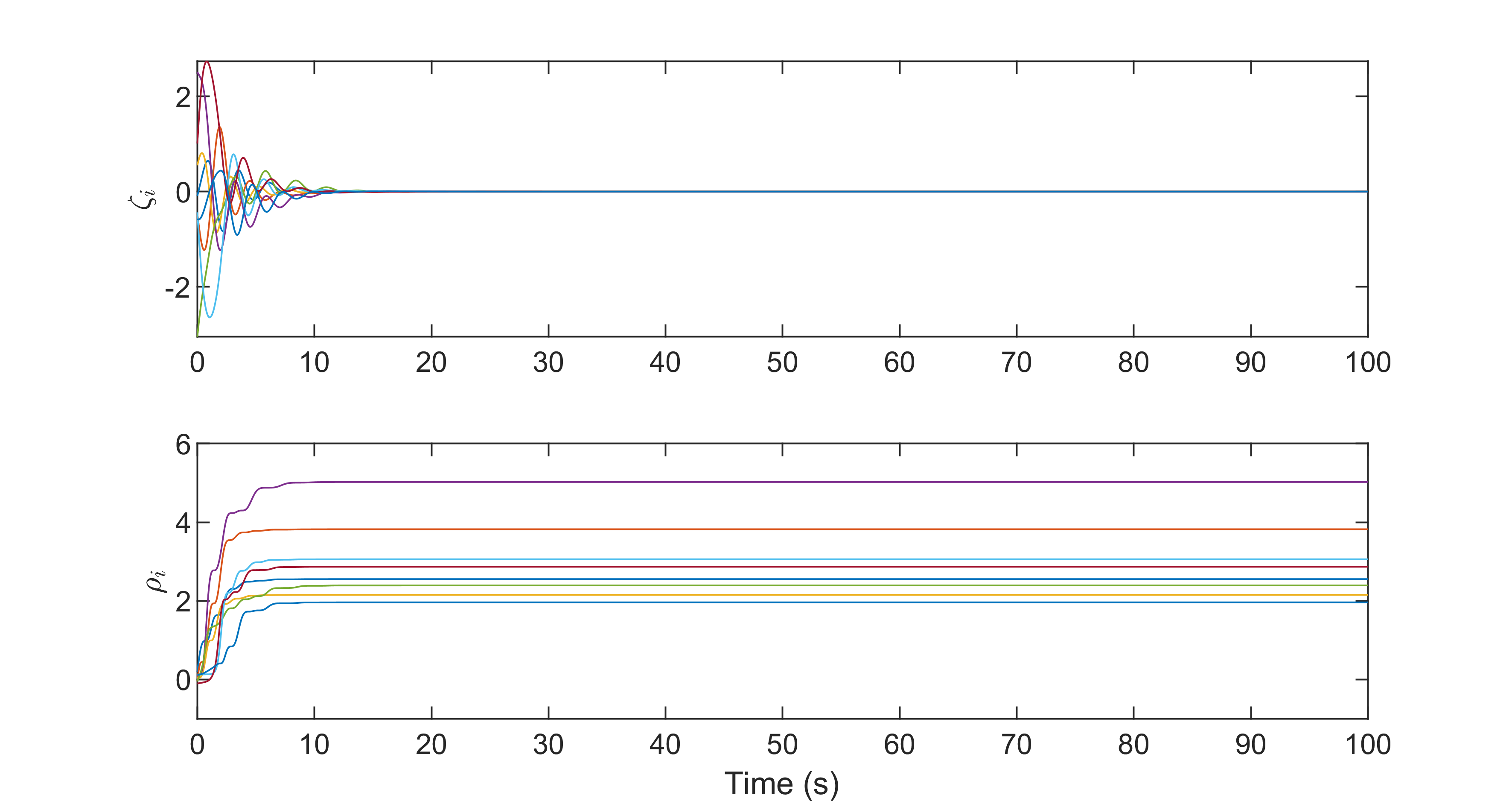} \centering
  \caption{The network information $\zeta_{i}$ and the adaptive
    parameters $\rho_{i}$ for 8-node network without a spanning tree
    via adaptive collaborative protocol.}\label{8c_wsyn_zeta_rho1}
\end{figure}

\begin{figure}[ht]
  \includegraphics[width=8cm]{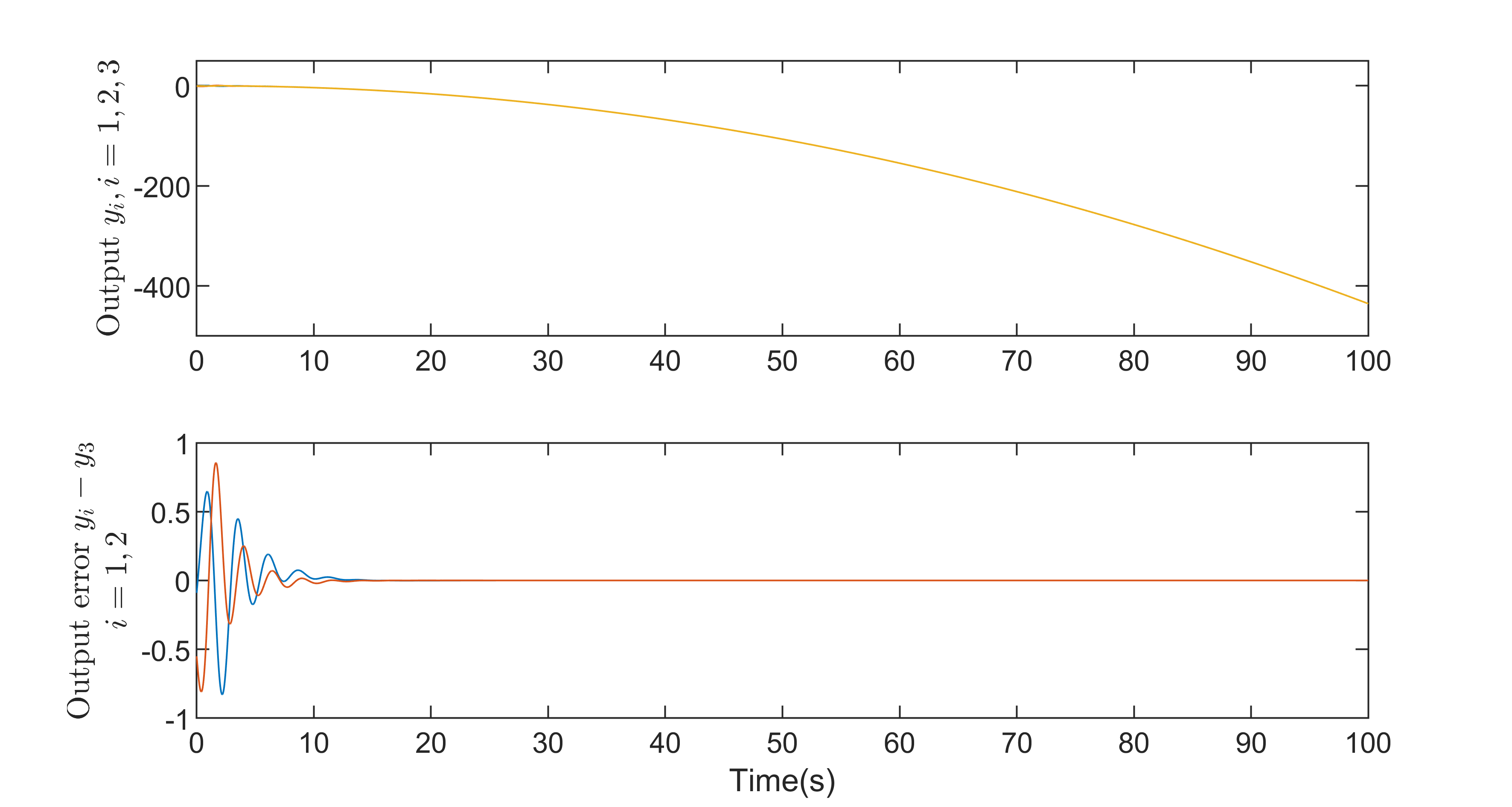} \centering
  \caption{The output synchronization and error output results of
    $y_i, i=1,2,3$ for 8-node network without a spanning tree via
    adaptive collaborative protocol.}\label{8c_wsyn_output123_error}
\end{figure}

\begin{figure}[ht]
  \includegraphics[width=8cm]{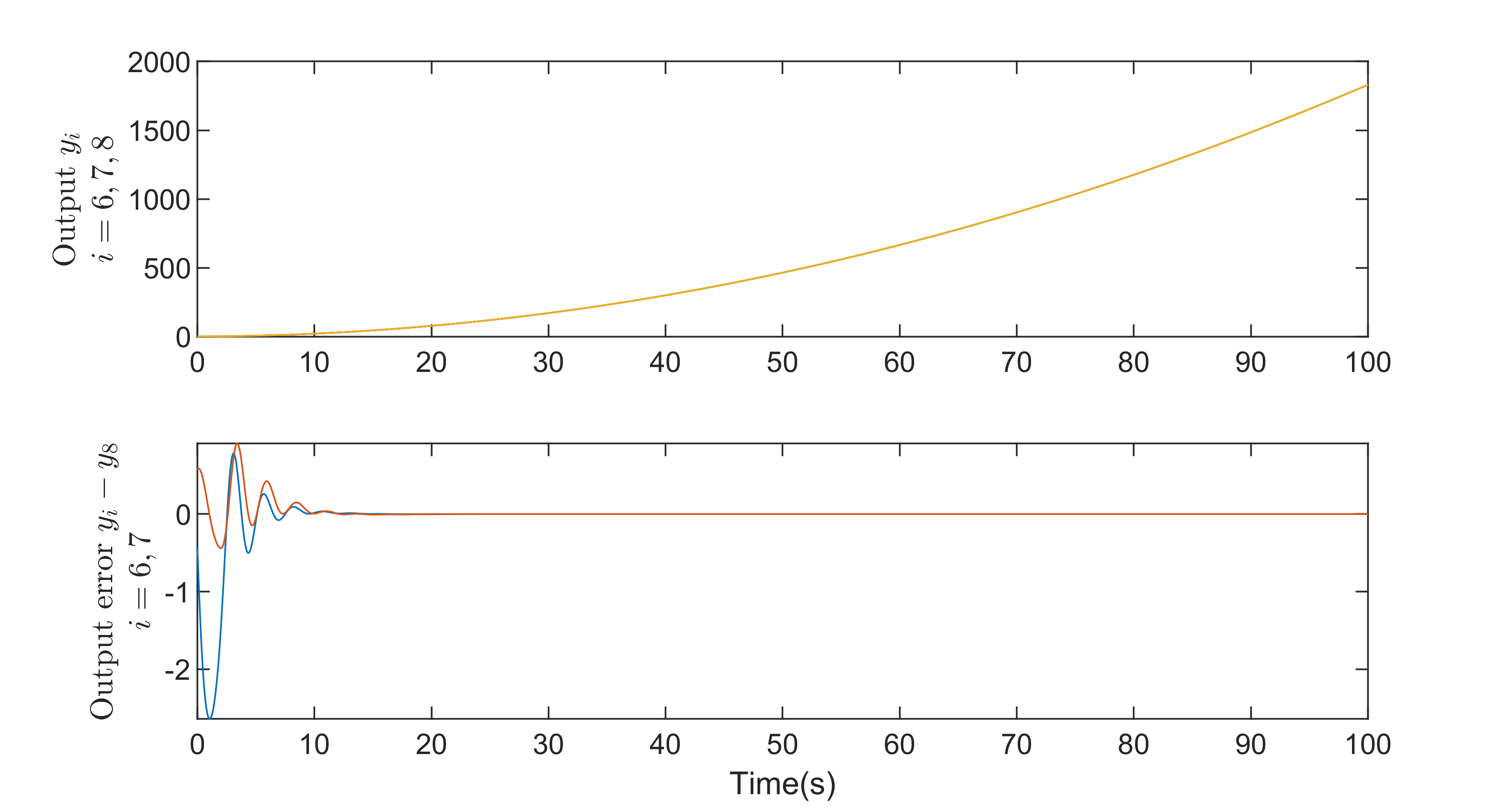} \centering
  \caption{The output synchronization and error output results of
    $y_i, i=6,7,8$ for 8-node network without a spanning tree via
    adaptive collaborative protocol.}\label{8c_wsyn_output678_error}
\end{figure}

For the network given in Figure \ref{f3}, weak output synchronization
is achieved by using the above adaptive collaborative protocol, see
Figure \ref{8c_wsyn_zeta_rho1}. Within Clusters 1 and 2 we achieve
output synchronization as illustrated in Figure
\ref{8c_wsyn_output123_error} and Figure 
\vref{8c_wsyn_output678_error}.

\begin{figure}[ht]
  \includegraphics[width=8cm]{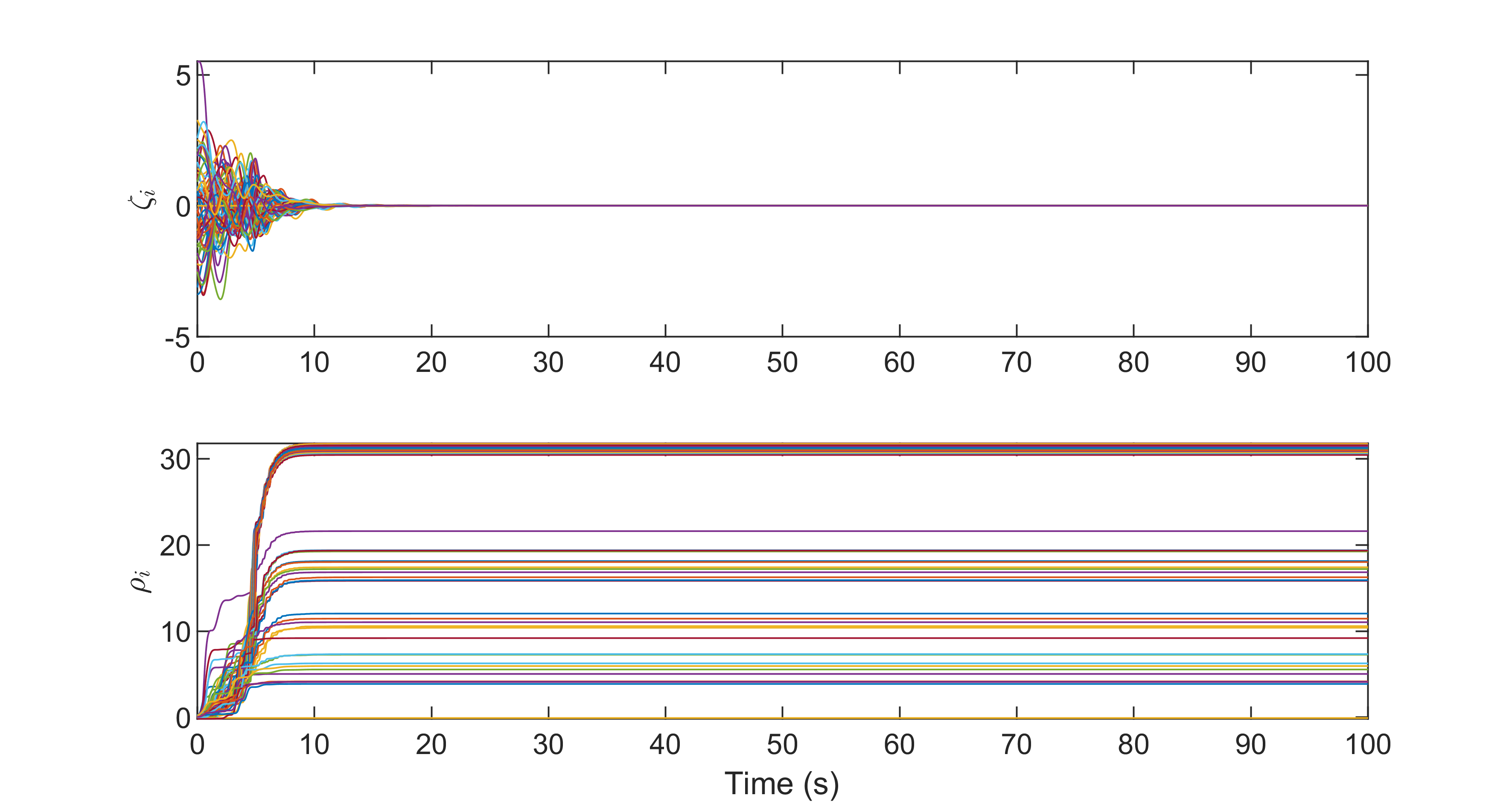} \centering
  \caption{The network information $\zeta_{i}$ and the adaptive
    parameters $\rho_{i}$ for 60-node network without a spanning tree
    via adaptive collaborative protocol.}\label{60c_wsyn_zeta_rho1}
\end{figure}

\begin{figure}[ht]
  \includegraphics[width=8cm]{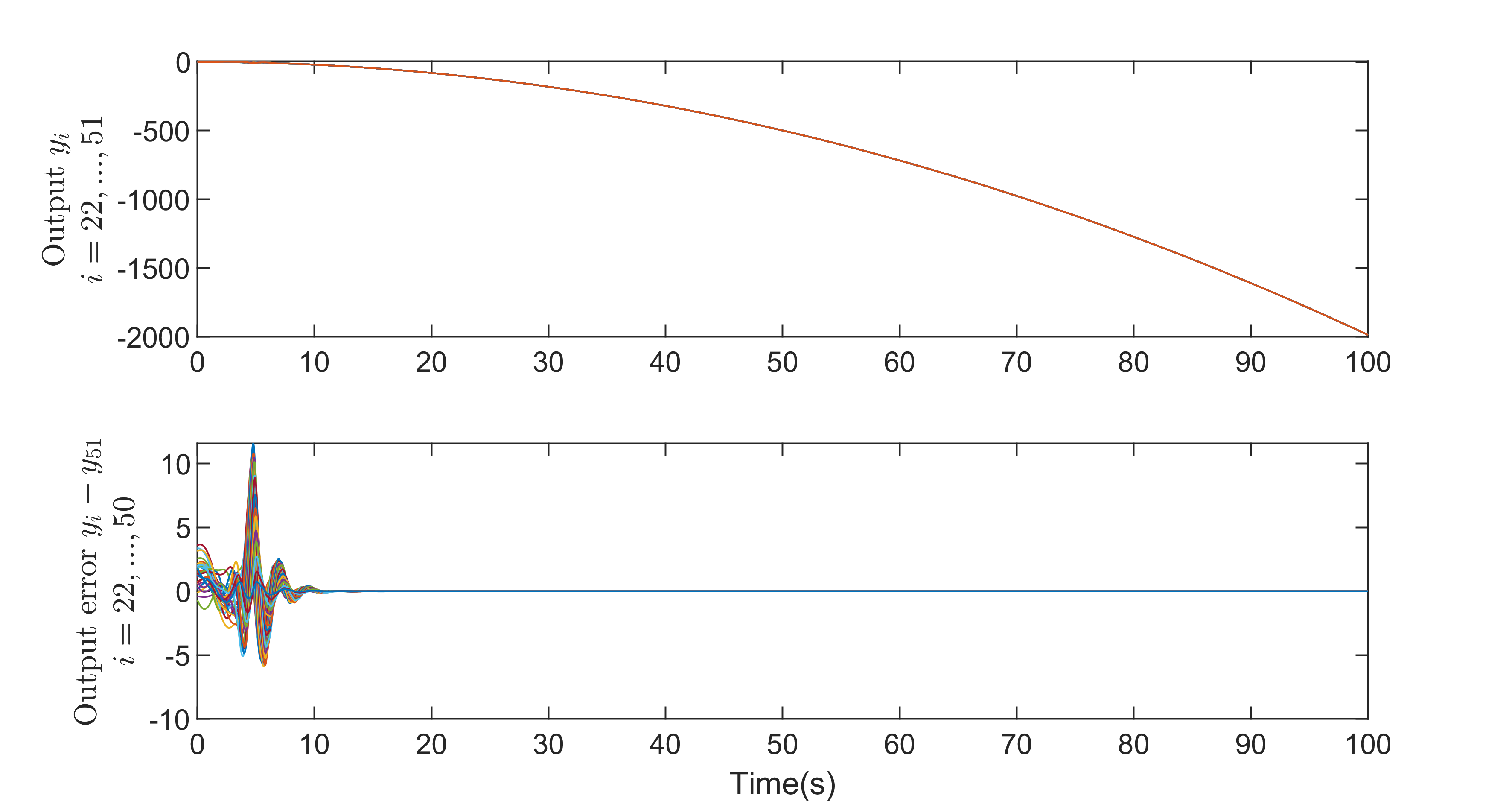}
  \centering
  \caption{The output synchronization and error state results of
    $y_i, i=22,\ldots,51$ for 60-node network without a spanning tree
    via adaptive collaborative
    protocol.}\label{60c_wsyn_output22-51_error}
\end{figure}

\begin{figure}[ht]
  \includegraphics[width=8cm]{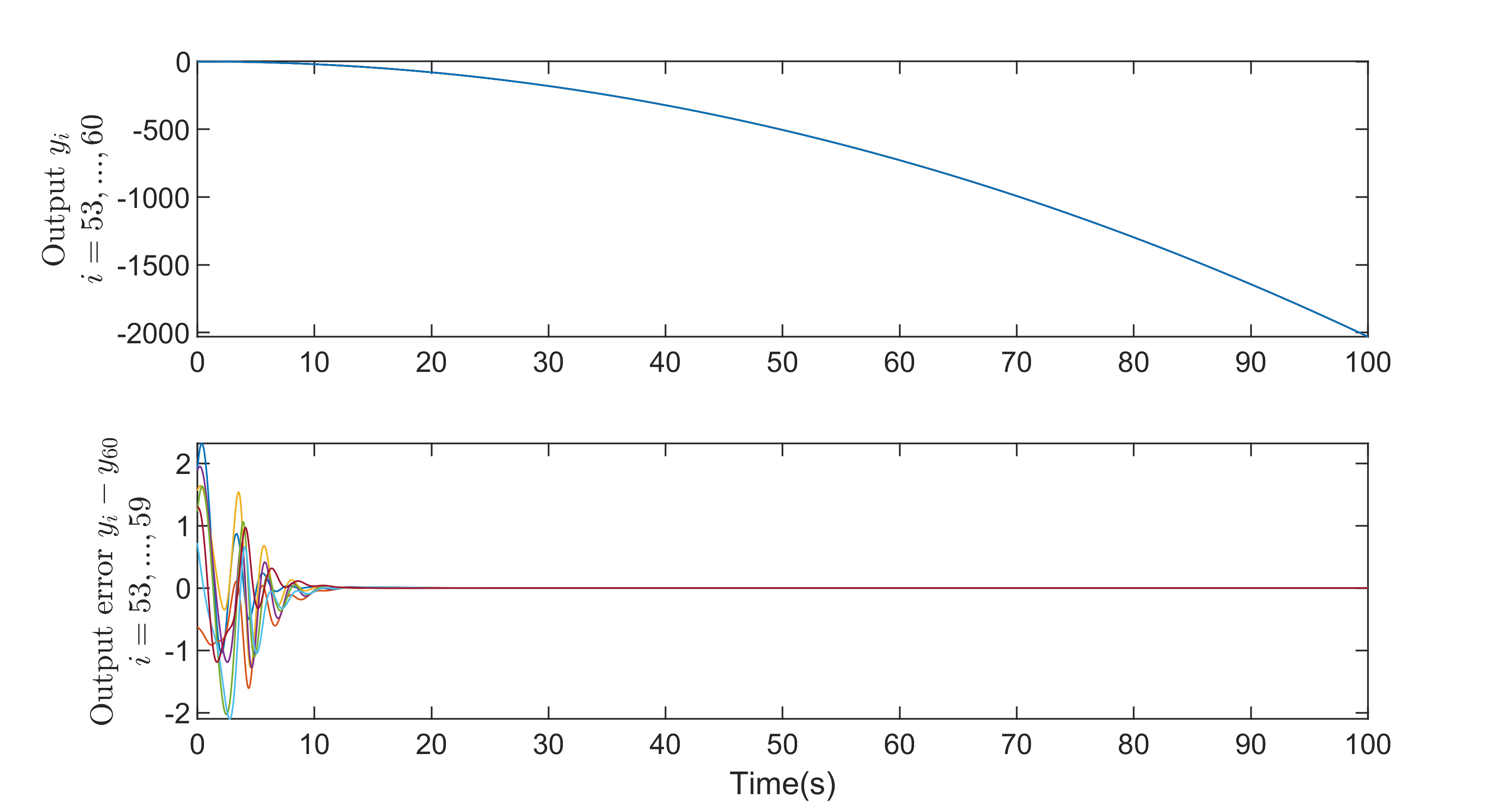}
  \centering
  \caption{The output synchronization and error output results of
    $y_i, i=53,\ldots,60$ for 60-node network without a spanning tree
    via adaptive collaborative
    protocol.}\label{60c_wsyn_output53-60_error}
\end{figure}

On the other hand, for the network shown in Figure \ref{f4}, we also
achieve weak output synchronization by the same adaptive collaborative
protocol, see Figure \vref{60c_wsyn_zeta_rho1}. Within Clusters 2 and 3
we achieve output synchronization, see Figures
\ref{60c_wsyn_output22-51_error} and \vref{60c_wsyn_output53-60_error}.

\FloatBarrier 

\section{Conclusion}

In this paper, we have designed a nonlinear adaptive protocol for MAS
with partial-state coupling to achieve output synchronization. When
the network includes a spanning tree, classical output synchronization
was achieved by designing scale-free non-collaborative and
collaborative adaptive protocols. If there is a fault in the network
to make the network no longer containing a directed spanning tree,
then we achieve weak output synchronization, a weaker synchronization
than the classical one. We have seen that the protocols guarantee a
stable response to these faults: within basic bicomponents we still
achieve synchronization and the outputs of the agents not contained in
a basic bicomponent converge to a convex combination of the asymptotic
behavior achieved in the basic bicomponents. Furthermore, the adaptive
protocol does not impose restrictions on the system poles which were
needed in linear scale-free design in
\cite{stoorvogel-saberi-liu-weak}.

\newcounter{equation2}
\setcounter{equation2}{\value{equation}}
\appendix
\setcounter{equation}{\value{equation2}}%

\section*{Appendix}

We recall the following lemmas from
\cite{donya-liu-saberi-stoorvogel-arxiv-delta} (with minor modifications):

\begin{lemma}\label{2.8}
  Consider a directed graph with Laplacian matrix $L$ which is
  strongly connected. Then, there exists $h_1,\ldots,h_N>0$ such that
  \begin{equation}\label{Hlyap}
    H^NL +L\T H^N \geq 2\gamma L\T L, 
  \end{equation}
  for some $\gamma >0$ with $H^N$ given by \eqref{HN} for $k=N$.
\end{lemma}

\begin{lemma}\label{2.9}
  The quadratic form
  \[
    V=z\T Q_{\rho} z
  \]
  with $Q_{\rho}$ given by \eqref{Qrho} is decreasing in $\rho_i$ for
  $i=1,\ldots, N$.
\end{lemma}

\bibliographystyle{elsarticle-num}
\bibliography{referenc}

\begin{thebibliography}{10}
\expandafter\ifx\csname url\endcsname\relax
  \def\url#1{\texttt{#1}}\fi
\expandafter\ifx\csname urlprefix\endcsname\relax\def\urlprefix{URL }\fi
\expandafter\ifx\csname href\endcsname\relax
  \def\href#1#2{#2} \def\path#1{#1}\fi

\bibitem{ren-atkins}
W.~Ren, E.~Atkins, Distributed multi-vehicle coordinate control via local
  information, Int. J. Robust \& Nonlinear Control 17~(10-11) (2007)
  1002--1033.

\bibitem{saber-murray2}
R.~Olfati-Saber, R.~Murray, Consensus problems in networks of agents with
  switching topology and time-delays, IEEE Trans. Aut. Contr. 49~(9) (2004)
  1520--1533.

\bibitem{wu-chua2}
C.~Wu, L.~Chua, Application of {K}ronecker products to the analysis of systems
  with uniform linear coupling, IEEE Trans. Circ. \& Syst.-I Fundamental theory
  and applications 42~(10) (1995) 775--778.

\bibitem{siljak}
D.~Siljak, Decentralized control of complex systems, Academic Press, London,
  1991.

\bibitem{corfmat-morse2}
J.~Corfmat, A.~Morse, Decentralized control of linear multivariable systems,
  Automatica 12~(5) (1976) 479--495.

\bibitem{li-wen-duan-ren-tac-2015}
Z.~Li, G.~Wen, Z.~Duan, W.~Ren, Designing fully distributed consensus protocols
  for linear multi-agent systems with directed graphs, IEEE Trans. Aut. Contr.
  60~(4) (2015) 1152--1157.

\bibitem{lv-li-ren-duan-chen-auto-2016}
Y.~Lv, Z.~Li, Z.~Duan, J.~Chen, Distributed adaptive output feedback consensus
  protocols for linear systems on directed graphs with a leader of bounded
  input, Automatica 74 (2016) 308--314.

\bibitem{liu-zhang-saberi-stoorvogel-auto}
Z.~Liu, M.~Zhang, A.~Saberi, A.~A. Stoorvogel, State synchronization of
  multi-agent systems via static or adaptive nonlinear dynamic protocols,
  Automatica 95 (2018) 316--327.

\bibitem{lv-li-scis-2021}
Y.~Lv, Z.~Li, Is fully distributed adaptive protocol applicable to graphs
  containing a directed spanning tree?, Science China Information Sciences
  65~(8) (2022) 189203 (1--2).

\bibitem{stoorvogel-saberi-liu-weak}
A.~A. Stoorvogel, A.~Saberi, Z.~Liu, Weak synchronization in heterogeneous
  multi-agent systems, https://arxiv.org/abs/2411.13806 (2025).

\bibitem{stoorvogel-saberi-liu2}
A.~Stoorvogel, A.~Saberi, Z.~Liu, Fault-tolerant properties of scale-free
  linear protocols for synchronization of homogeneous multi-agent systems,
  http://arxiv.org/abs/2403.18200 (2023).

\bibitem{royle-godsil}
C.~Godsil, G.~Royle, Algebraic graph theory, Vol. 207 of Graduate Texts in
  Mathematics, Springer-Verlag, New York, 2001.

\bibitem{ren-book}
W.~Ren, Y.~Cao, Distributed coordination of multi-agent networks,
  Communications and Control Engineering, Springer-Verlag, London, 2011.

\bibitem{wu-book}
C.~Wu, Synchronization in complex networks of nonlinear dynamical systems,
  World Scientific Publishing Company, Singapore, 2007.

\bibitem{li-duan-chen-huang}
Z.~Li, Z.~Duan, G.~Chen, L.~Huang, Consensus of multi-agent systems and
  synchronization of complex networks: A unified viewpoint, IEEE Trans. Circ.
  \& Syst.-I Regular papers 57~(1) (2010) 213--224.

\bibitem{stoorvogel-saberi-liu}
A.~Stoorvogel, A.~Saberi, Z.~Liu, The role of local bounds on neighborhoods in
  the network for scale-free state synchronization of multi-agent systems, Int.
  J. Robust \& Nonlinear Control 34~(15) (2024) 10555--10570.

\bibitem{liu-saberi-stoorvogel-tac-2023}
Z.~Liu, A.~Saberi, A.~Stoorvogel, Scale-free non-collaborative linear protocol
  design for a class of homogeneous multi-agent systems, IEEE Trans. Aut.
  Contr. 69~(5) (2024) 3333--3340.

\bibitem{stoorvogel-saberi-liu-masoumi}
A.~A. Stoorvogel, A.~Saberi, Z.~Liu, Z.~Masoumi, Weak state synchronization of
  homogeneous multi-agent systems with adaptive protocols, in preparation
  (2025).

\bibitem{qu-book-2009}
Z.~Qu, Cooperative control of dynamical systems: applications to autonomous
  vehicles, Spinger-Verlag, London, U.K., 2009.

\bibitem{donya-liu-saberi-stoorvogel-arxiv-delta}
D.~Nojavanzadeh, Z.~Liu, A.~Saberi, A.~A. Stoorvogel, Scalable $\delta$-level
  coherent state synchronization of multi-agent systems with adaptiveprotocols
  and bounded disturbances, Int. J. of Adaptive Control and Signal Processing
  39~(3) (2025) 497--516.

\end{thebibliography}
\end{document}